# CraftSPH: A composable framework for differentiable smoothed particle hydrodynamics implemented in PyTorch

Authors: Gen Matono[a*], Shujiro Fujioka[b], and Mayuko Nishio[c]

[a] Graduate student, Graduate School of Science and Technology, University of Tsukuba (ORCID number: 0009-0004-0069-2120)

[b] Graduate Student, Department of Civil Engineering, Kyushu University (ORCID number: 0009-0000-6414-4909)

[c] Ph.D., Associate professor, Institute of Systems and Information Engineering, University of Tsukuba (ORCID number: 0000-0003-1079-2577)

## ABSTRACT

Smoothed particle hydrodynamics (SPH) is well suited to large-deformation fluid problems, and the growing use of automatic differentiation in engineering applications has motivated interest in differentiable SPH solvers. In this study, a composable differentiable SPH framework, CraftSPH, has been developed. In CraftSPH, major computational operations are implemented as independent modules that users can combine to construct solvers for different purposes. This design supports high-accuracy discretization and both explicit and implicit computation within a common framework, while allowing individual components to be maintained and extended. Selected numerical operations also support automatic differentiation (AD) for physical-parameter estimation and potential coupling with deep learning models. These capabilities were demonstrated through forward and inverse analyses. Forward analyses using an implicit SPH scheme with high-accuracy discretization showed good agreement with analytical and reference solutions for Poiseuille flow and a rising bubble, while capturing the key experimental trends in two- and three-dimensional dam-break problems. In inverse analyses, explicit and implicit SPH schemes were used to estimate kinematic viscosity in Taylor–Green vortex flow and Reynolds number in lid-driven cavity flow. The results also showed that the choice of operations included in the computational graph affects the computed gradients. Together, these findings demonstrate that CraftSPH enables users to assemble problem-specific solvers for forward and inverse analyses within a common framework. The code is publicly available at https://github.com/GenMNL/CraftSPH

# 1. INTRODUCTION

Smoothed particle hydrodynamics (SPH) [1][2] is a representative particle method that discretizes a continuum into a set of particles and tracks their motion. Since SPH does not require complex computational meshes, it is suitable for problems involving large deformation. However, the movement of computational particles can lead to nonuniform particle distributions, which have traditionally posed a challenge to maintaining the accuracy of differential approximations [3][4]. To address this limitation, numerous studies have proposed high-accuracy differential approximation models [5] and various stabilization techniques such as density diffusion [6] and particle shifting techniques [7]. In recent years, practical levels of accuracy and stability have been achieved by appropriately combining these numerical schemes. Owing to these significant research advances, SPH has been applied to a wide range of fields, including fluid analysis[8], geotechnical analysis [9], and biological tissue analysis [10], and further improvements in accuracy and expansion of its applications are expected.

In recent years, numerical solvers have increasingly been used not only for forward analysis but also for inverse analysis and as components of deep learning models. For example, inverse problems that estimate physical parameters, such as material properties, from observational data [11], and neural networks that incorporate numerical computations [12][13] require gradients of the numerical results with respect to model parameters. For such applications, differentiable solvers based on differentiable programming provide a promising approach for efficiently obtaining the required gradients through automatic differentiation (AD) [14]. Furthermore, evaluating the effectiveness of a learned deep learning model requires comparison with existing numerical solvers in terms of both computational time and accuracy. In this context, the importance of implementing deep learning models and reference numerical solvers in the same programming language and computational environment has been indicated to enable a fair comparison in terms of both computational performance and accuracy [15]. Therefore, an evaluation environment is required in which both the deep learning model and the numerical solver are implemented using the same programming language and computational framework, thereby enabling their speed and accuracy to be assessed under consistent conditions. These requirements have increased the importance of differentiable solvers implemented using frameworks such as PyTorch [16], JAX [17], and TensorFlow [18], which provide native support for deep learning models and AD.

Against this background, differentiable solvers have been developed for various numerical methods, including DiffTaichi[19], JAX-MD[20], JAX-FEM[21], AD-LBM[22], PICT[23], and JAX-MPM[24], and have been used for inverse analysis based on AD and learning-based physical simulation. Similar developments are being observed in the field of SPH. Warp[25] and $\mathbf{\Phi}_{\mathrm{Flow}}$[26] are general-purpose frameworks with limited support for SPH. In contrast, DiffFR[27] and SPNet[11] are SPH models designed for specific tasks, while JAX-SPH[28] and diffSPH[29] provide dedicated support for differentiable SPH solvers. Owing to their high computational performance and compatibility with AD, these solvers have made significant contributions to the development of differentiable particle methods. To build on these advances and facilitate the use of differentiable SPH solvers for different research problems, numerical methods should be adaptable to the problem at hand. In particular, it is useful to be able to select and combine spatial discretization, pressure computation schemes, boundary condition treatments, and stabilization methods. Organizing these operations as independent components also facilitates the construction and maintenance of customized numerical schemes, including those that incorporate high-accuracy SPH discretization and semi-implicit pressure computation. For applications involving inverse analysis or integration with deep learning models, the resulting solver must also make clear which numerical operations are included in AD.

In this study, we develop CraftSPH, a composable differentiable SPH framework designed to satisfy the requirements described above. In CraftSPH, key components of SPH analysis, including neighbor search, kernel functions, differential operators, boundary condition handling, and stabilization techniques, are implemented as independent modules. The framework is designed on the premise that users construct their own SPH solvers by intuitively combining these modules, thereby enabling the incorporation and extension of numerical schemes, including high-accuracy discretization and implicit computation. Additionally, major particle data, including physical variables and material properties, are managed within a unified particle object, enabling the easy addition of new physical quantities and numerical modules. By implementing these components in PyTorch, the framework supports the construction of hybrid models combining SPH and deep learning and gradient computation for inverse analysis and deep learning model training on both CPUs and GPUs. Moreover, implementing both SPH solvers and deep learning models in PyTorch supports comparisons under more consistent implementation and execution conditions.

This paper presents the design principles and major components of CraftSPH and evaluates its effectiveness for both forward and inverse analyses, with a particular focus on fluid dynamics. The forward analysis demonstrates that advanced SPH solvers can be constructed by flexibly combining the proposed modules, whereas the inverse analysis evaluates the use of AD for gradient computation. Our previous study [30] used CraftSPH with deep learning models and for comparative evaluations; these are prior applications rather than experiments conducted in this paper. The remainder of this paper is organized as follows. Section 2 describes the governing equations and mathematical formulations of SPH discretization used in the present fluid-dynamics validations. Section 3 presents the implementation details of CraftSPH, including data management, computational modules, and solver construction. Sections 4 and 5 evaluate the framework through forward and inverse analyses, respectively. Finally, Section 6 summarizes the main findings and future developments.

**Contributions.** Our main contributions are as follows:

- **A composable SPH framework:** Shared particle data and independent numerical components allow users to define solver procedures and maintain or extend individual components.
- **Support for high-accuracy and implicit SPH computation:** High-accuracy differential operators, stabilization techniques, and matrix-form operators for explicit and implicit solver construction.
- **Forward analysis application:** Four distinct fluid-flow problems demonstrate the effectiveness of assembling problem-specific solvers by flexibly combining common components.
- **Inverse analysis application:** Two parameter-estimation problems demonstrate the applicability of AD and investigate how the choice of differentiation paths affects the computed gradients.

## 2. DETAIL OF SPH METHOD

This section describes the governing equations and the spatial and temporal discretization, with a focus on the incompressible fluid flows considered in the present validation. However, CraftSPH is a general-purpose differentiable solver framework for SPH, in which users construct their own solvers or integrate SPH computations into deep learning models by assembling the available components. Therefore, although this section presents the governing equations used in the present validation, solvers for other governing equations can also be constructed flexibly by selecting and combining the components in different ways.

## 2.1. Governing equation for fluid dynamics

In the numerical analysis of incompressible fluid flows using the SPH method, the governing equations are the Navier-Stokes equation in the Lagrangian description and the continuity equation, which are described as follows:

$$\frac{D\boldsymbol{v}}{Dt} = -\frac{1}{\rho}\nabla p + \nu\nabla^2\boldsymbol{v} + \frac{1}{\rho}\boldsymbol{f}_{\text{ext}} \tag{1}$$

and

$$\frac{D\rho}{Dt} + \rho\nabla\cdot\boldsymbol{v} = 0 \tag{2}$$

where $t$ is time, $\boldsymbol{v}$ is velocity vector, $p$ is pressure, $\rho$ is density, $\nu$ is kinematic viscosity, and $\boldsymbol{f}_{\text{ext}}$ is the external force, including effects such as gravity force $\rho\boldsymbol{g}$ and surface tension force $\boldsymbol{f}_{\sigma}$.

## 2.2. Spatial discretization in SPH method

CraftSPH supports three spatial differential approximation models. The first is classical SPH [1][2][31][32], which provides the most basic formulation. The second is corrected SPH (CSPH) [33][34], which offers improved accuracy and stability. The third is least squares SPH (LSSPH) [5], which generalizes these formulations to achieve higher-order accuracy. These models are implemented as independent differential operator modules, allowing users to select an appropriate discretization model according to the target problem and the required accuracy. The model can be readily selected through the configuration file. In addition, newly developed state-of-the-art models can be incorporated by implementing their formulations and adding them to the available model options. This section describes the mathematical background of these differential approximation models.

### 2.2.1. Classical SPH

In the classical SPH, physical quantity $\phi_i$ of arbitrary particle $i$ is approximated as follows [1][2]:

$$\phi_i \approx \sum_{j\in\mathbb{P}_i} V_j\phi_j W_{ij} \tag{3}$$

where $i$ and $j$ are index of a main particle and its neighbor particles, $\mathbb{P}_i$ is a neighbor particle sets of a particle $i$, $V_i$ is volume of particle $i$ and $W_{ij}$ is a kernel function defined between relative position vector $\boldsymbol{r}_{ij} \coloneqq \boldsymbol{x}_j - \boldsymbol{x}_i$ ($\boldsymbol{x} \in \mathbb{R}^d$ is coordinate). The kernel function $W_{ij} \coloneqq W(\boldsymbol{r}_{ij}; h)$ is a smooth function that determines the degree of influence between particle $i$ and its neighboring particles, where $h$ denotes the smoothing length defines the spatial extent of the kernel support. Using the gradient of the kernel function, the SPH approximations of the differential operators are expressed as follows:

$$\nabla\phi_i \approx \sum_{j\in\mathbb{P}_i} V_j \phi_{ij} \nabla W_{ij} \tag{4}$$

$$\nabla^2\phi_i \approx 2\sum_{j\in\mathbb{P}_i} V_j \frac{\phi_{ij}}{\left|\boldsymbol{r}_{ij}\right|^2} \boldsymbol{r}_{ij} \cdot \nabla W_{ij} \tag{5}$$

Here, $\phi_{ij} \coloneqq \phi_j - \phi_i$ is difference in the physical quantity. Eq. (4) was originally introduced for the approximation of divergence [1]. Although an alternative formulation using $\phi_j$ instead of $\phi_{ij}$ is also available, the formulation given in Eq. (4) is more commonly used in fluid flow simulations. On the other hand, various classical formulations have also been proposed for second-order derivatives, among which Eq. (5) is one of the most widely used approximations of the Laplacian operator.

Classical SPH has a simple computational procedure. However, its approximation accuracy depends strongly on the particle arrangement and deteriorates when the particle distribution becomes nonuniform [3][4]. More specifically, for irregular particle distributions, the gradient approximation exhibits zeroth-order convergence, whereas the Laplacian approximation exhibits a negative convergence. Because computational particles in SPH move with the fluid motion, disturbances in the particle distribution frequently occur during simulations. Depending on the problem, this limitation can therefore lead to critical numerical errors. The CSPH and LSSPH formulations described below are effective in addressing this issue.

### 2.2.2. Corrected SPH (CSPH)

To achieve high-precision SPH analysis, CSPH first defines the moment matrix $\boldsymbol{L}_i^{\mathrm{CSPH}}$ as follows:

$$\boldsymbol{L}_i^{\mathrm{CSPH}} := \sum_{j \in \mathbb{P}_i} V_j \nabla W_{ij} \otimes \boldsymbol{r}_{ij} \tag{6}$$

where $\otimes$ denotes the outer product, defined for two vectors $\boldsymbol{a}$ and $\boldsymbol{b}$ as $\boldsymbol{a} \otimes \boldsymbol{b} = \boldsymbol{a}\boldsymbol{b}^{\mathrm{T}}$. If the moment matrix is nonsingular, its inverse is uniquely determined. Using this inverse, the corrected kernel gradient $\widetilde{\nabla} W_{ij}$ is defined as follows [33][34]:

$$\widetilde{\nabla} W_{ij} := \left(\boldsymbol{L}_i^{\mathrm{CSPH}}\right)^{-1} \nabla W_{ij} \tag{7}$$

When the particles are arranged on a regular lattice, the moment matrix becomes the identity matrix, and the corrected kernel gradient coincides with the original kernel gradient.

This correction improves the convergence of the gradient approximation for irregular particle distributions, allowing first-order convergence to be achieved even for disordered particle arrangements. In practical computations, the kernel gradient in Eq. (4) is replaced by the corrected kernel gradient when evaluating the differential operator. However, for irregular particle distributions, this correction does not improve convergence beyond first order, nor does it improve the convergence of the Laplacian approximation. Therefore, when higher accuracy is required or when the method is applied to viscosity-dominated problems, a formulation capable of accurately approximating derivatives of arbitrary order accuracy is desirable. In addition, the moment matrix may become singular near free surfaces, where the number of neighboring particles is limited, or when the particle distribution is strongly biased. To address this issue, approaches such as switching to eliminating off-diagonal components [35], a lower-order accuracy differential model near free surfaces [36], and applying weighted kernel gradient correction [37] have been proposed.

### 2.2.3. Least squares SPH (LSSPH)

LSSPH provides a more general polynomial-reconstruction framework for high-accuracy differential approximation. In principle, it can approximate derivatives of arbitrary order with arbitrary accuracy. In LSSPH, the distribution of a physical quantity around a particle is represented by a Taylor expansion, and its derivative coefficients are determined using the moving least-squares method. First, the physical quantity $\phi_j$ around particle $i$ is expanded using a Taylor series as follows:

$$\phi_j \approx \boldsymbol{P}_{ij}^p \cdot \boldsymbol{\delta}^p \phi_i \tag{8}$$

where, $\boldsymbol{P}_{ij}^p$ is the polynomial basis vector containing terms from zeroth to $p$th-order, and $\boldsymbol{\delta}^p$ is the differential operator vector containing derivatives from zeroth to $p$th-order. Let $\boldsymbol{\delta}^p \phi_i$ be treated as the unknown variable $\boldsymbol{X}_i$, and consider weighted least-squares error $J(\boldsymbol{X}_i)$ as

$$J(\boldsymbol{X}_i) \coloneqq \sum_{j \in \mathbb{P}_i} \mathrm{V_j} W_{ij} \left(\phi_j - \boldsymbol{P}_{ij}^p \cdot \boldsymbol{X}_i\right)^2 \tag{9}$$

The normal equations for minimizing $J(\boldsymbol{X}_i)$ are given by

$$\boldsymbol{L}_i^{\mathrm{LSSPH}} \coloneqq \sum_{j \in \mathbb{P}_i} V_j W_{ij} \boldsymbol{P}_{ij}^p \otimes \boldsymbol{P}_{ij}^p \tag{10}$$

$$\boldsymbol{L}_i^{\mathrm{LSSPH}} \boldsymbol{X}_i = \sum_{j \in \mathbb{P}_i} V_j W_{ij} \boldsymbol{P}_{ij}^p \phi_j \tag{11}$$

Here, $\boldsymbol{L}_i^{\mathrm{LSSPH}}$ is a moment matrix for LSSPH. When the moment matrix $\boldsymbol{L}_i^{\mathrm{LSSPH}}$ is nonsingular, its inverse is uniquely determined, and consequently, $\boldsymbol{X}_i$ is uniquely defined.

$$\boldsymbol{X}_i = \left(\boldsymbol{L}_i^{\mathrm{LSSPH}}\right)^{-1} \sum_{j \in \mathbb{P}_i} V_j W_{ij} \boldsymbol{P}_{ij}^p \phi_j \tag{12}$$

The resulting unknown vector $\boldsymbol{X}_i$ corresponds to $\boldsymbol{\delta}^p \phi_i$, which contains the approximated derivatives up to $p$th-order. A closely related formulation has also been introduced in the LSMPS [38], where the formulation corresponding to Eqs. (10)-(12) is classified into Type-A and Type-B depending on whether the zeroth-order term is treated as an unknown. In Type-A, the physical quantity $\phi_i$ carried by particle $i$ is treated as a known value. Let $\boldsymbol{P}_{ij}^{1:p}$ be defined as the polynomial basis vector without the constant term. The derivative vector $\boldsymbol{\delta}^{1:p} \phi_i$, which contains the approximated derivatives from first to $p$th-order, is then obtained as follows:

$$\boldsymbol{\delta}^{1:p}\phi_i = \left(\boldsymbol{L}_{i,\mathrm{A}}^{\mathrm{LSSPH}}\right)^{-1} \sum_{j\in\mathbb{P}_i} V_j W_{ij}\, \boldsymbol{P}_{ij}^{1:p}(\phi_j - \phi_i) \tag{13}$$

Here, $\boldsymbol{L}_{i,\mathrm{A}}^{\mathrm{LSSPH}}$ is obtained from Eq. (10) by setting $\boldsymbol{P}_{ij}^{p} = \boldsymbol{P}_{ij}^{1:p}$. On the other hand, Type-B treats the zeroth-order reconstructed value as an unknown, together with the spatial derivatives up to order $p$. Let $\boldsymbol{P}_{ij}^{0:p}$ be defined as the polynomial basis vector including the constant term. The derivative vector $\boldsymbol{\delta}^{0:p}\phi_i$, which contains the approximated derivatives from zeroth to $p$th-order, is then obtained as follows:

$$\boldsymbol{\delta}^{0:p}\phi_i = \left(\boldsymbol{L}_{i,\mathrm{B}}^{\mathrm{LSSPH}}\right)^{-1} \sum_{j\in\mathbb{P}_i} V_j W_{ij}\, \boldsymbol{P}_{ij}^{0:p}\phi_j \tag{14}$$

Here, $\boldsymbol{L}_{i,\mathrm{B}}^{\mathrm{LSSPH}}$ is obtained from Eq. (10) by setting $\boldsymbol{P}_{ij}^{p} = \boldsymbol{P}_{ij}^{0:p}$. In practical applications, Type-A is more commonly used because of its smaller moment matrix and lower computational cost. Type-B is used when a higher-order accuracy of the zeroth-order term is required.

Following these formulations, derivatives of arbitrary order can be approximated with arbitrary accuracy. However, practical applications are limited by the number of neighboring particles and numerical stability. Therefore, second-order accuracy models are commonly used for gradient approximation and first-order accuracy models for the Laplacian under distorted particle distributions. Even high-order accuracy models can be less stable than low-order accuracy models because of issues such as the loss of pairwise conservation and missing neighboring particles. This problem becomes more pronounced for LSSPH formulations, which require a larger number of neighboring particles than CSPH and are therefore more susceptible to numerical instability. For this reason, stabilization treatments such as regularization of unstable particle configurations [39][40][41] and switching between differential approximation models [42] are often required for robust simulations involving strongly distorted particle distributions or truncated neighborhoods.

### 2.3. Temporal discretization of SPH method

Temporal discretization in numerical analysis can be broadly classified into explicit and implicit approaches. Explicit methods have an advantage for relatively simple computational procedures and low cost per time step because the state at the next time step is directly computed from the state at the current time step. However, the allowable time-step size is often severely restricted by numerical stability. In contrast, implicit methods require the iterative calculation to solve unknown quantities at the next time step, which increases the computational cost per time step. Nevertheless, they generally allow larger time-step sizes than explicit methods and can therefore reduce the total computational cost depending on the problem.

On the other hand, most existing differentiable SPH solvers [28][29] have been developed mainly for explicit methods. When considering the use of differentiable solvers for comparison with deep learning, explicit methods are appropriate for demonstrating speedups per time step. However, comparing them against implicit methods is crucial for demonstrating speedups achieved by using larger time steps. Particularly in fluid analysis, explicit methods suffer severe time-step constraints; to maintain near-constant density, an artificially large speed of sound must be introduced, which drastically limits the maximum allowable time step due to the Courant-Friedrichs-Lewy (CFL) condition [46]. In contrast, implicit methods overcome this speed of sound limitation by directly enforcing the incompressibility condition, thereby allowing for significantly larger time steps. Moreover, implicit methods are also important from the viewpoint of accuracy in incompressible flow problems because they can directly satisfy the incompressibility condition and suppress fluctuations in density and pressure [43]. From these reasons, it is crucial to be able to construct not only explicit methods but also implicit methods, considering both speed and accuracy. This section introduces two standard SPH approaches for incompressible fluids that are also used in the numerical experiments presented later: the explicit-type weakly compressible SPH (WCSPH) [44] and the implicit-type incompressible SPH (ISPH) [45].

#### 2.3.1. WCSPH

In the WCSPH method [44], the Navier-Stokes equation shown in eq. (1) are discretized as follows:

$$\frac{D\boldsymbol{v}_i}{Dt} = -\sum_{j \in \mathbb{P}_i} \left[ m_j \left( \frac{p_i}{\rho_i^2} + \frac{p_j}{\rho_j^2} + \Pi_{ij} \right) \right] \nabla W_{ij} + \frac{1}{\rho_i} \boldsymbol{f}_{\mathrm{ext},i} \tag{15}$$

where, $m$ is mass of each particle and $\Pi$ is viscosity term including artificial viscosity. Typically, the viscosity term is given by following equations:

$$\Pi_{ij} = \begin{cases} -\dfrac{\alpha C_s \mu_{ij} + \beta \mu_{ij}^2}{\bar{\rho}_{ij}}, & \boldsymbol{v}_{ij} \cdot \boldsymbol{r}_{ij} < 0 \\ 0, & \boldsymbol{v}_{ij} \cdot \boldsymbol{r}_{ij} \geq 0 \end{cases} \tag{16}$$

with

$$\mu_{ij} = \frac{h \boldsymbol{v}_{ij} \cdot \boldsymbol{r}_{ij}}{\left|\boldsymbol{r}_{ij}\right|^2 + 0.01h^2} \tag{17}$$

$$\bar{\rho}_{ij} = \frac{\rho_i + \rho_j}{2} \tag{18}$$

where, $C_s$ is speed of sound and $\alpha,\ \beta$ are parameters. In [44], the parameters are set to $\alpha = 0.01,\ \beta = 0$. The equations are solved by coupling the above expression with the continuity equation shown in Eq. (2). However, in the WCSPH framework, the system of equations is not closed with these two equations alone because density is also treated as a variable. Therefore, an equation of state, which relates pressure, density, and internal energy as in compressible fluid solvers, is required to determine the pressure. Typically, the Tait equation of state is employed as follows:

$$p_i = B\left(\left(\frac{\rho_i}{\rho^0}\right)^{\gamma} - 1\right) \tag{19}$$

where $B = C_s^2 \rho^0 / \gamma$ is the reference pressure, $\rho^0$ is initial density and $\gamma = 7$. These equations are solved sequentially through time integration. Specifically, the velocity is integrated using the explicit Euler method, while the particle positions are updated using the implicit Euler method. This combination constitutes the symplectic Euler method. In this way, the WCSPH method is an explicit scheme, where the solution at the next time step can be obtained solely from the current values. The main advantage of explicit schemes is their computational efficiency, as values can be updated sequentially, allowing for fast computation per time step. However, the time step is limited by the CFL condition [46], as described below.

$$\Delta t \leq C_{\mathrm{CFL}} \left( \frac{h}{C_s + \max\limits_i \|\boldsymbol{v}_i\|_2} \right) \tag{20}$$

where $C_{\mathrm{CFL}} < 1$ is Courant number. In WCSPH, an artificial speed of sound is introduced to restrict density variations to approximately 1%, which typically requires $C_s \geq 10\, V_{\mathrm{ref}}$, where $V_{\mathrm{ref}}$ is reference velocity. Consequently, $C_s$ is substantially larger than the characteristic flow velocity and the acoustic term usually imposes a severe restriction on the time-step size [47].

When WCSPH are used in inverse analysis in section 5, pressure is evaluated using the Tait equation of state, and velocity is updated using this pressure, as expressed by the following equations:

$$p^t = B\left(\left(\frac{\rho^t}{\rho^0}\right)^{\gamma} - 1\right) \tag{21}$$

$$\boldsymbol{v}^{t+1} = \boldsymbol{v}^t - \Delta t \left( \frac{1}{\rho^0} \nabla p^t - \nu \nabla^2 \boldsymbol{v}^t - \frac{1}{\rho^0} \boldsymbol{f}_{\mathrm{ext}} \right) \tag{22}$$

### 2.3.2. ISPH

In the ISPH method [45], the incompressibility approximation is applied by assuming that the fluid density remains constant in time. The governing equations are given as follows:

$$\frac{D\boldsymbol{v}}{Dt} = -\frac{1}{\rho^0} \nabla p + \nu \nabla^2 \boldsymbol{v} + \frac{1}{\rho^0} \boldsymbol{f}_{\mathrm{ext},i} \tag{23}$$

$$\nabla \cdot \boldsymbol{v} = 0 \tag{24}$$

In the ISPH method, time integration is performed based on the projection method [48], also known as the fractional step method, as described below:

$$\boldsymbol{v}^* = \boldsymbol{v}^t + \Delta t \left( \nu \nabla^2 \boldsymbol{v} + \frac{1}{\rho^0} \boldsymbol{f}_{\mathrm{ext}} \right) \tag{25}$$

$$\nabla \cdot \left(\frac{1}{\rho^0} \nabla p^{t+1}\right) = \frac{\nabla \cdot \boldsymbol{v}^*}{\Delta t} \tag{26}$$

$$\boldsymbol{v}^{t+1} = \boldsymbol{v}^* - \Delta t \left(\frac{1}{\rho^0} \nabla p^{t+1}\right) \tag{27}$$

where $\boldsymbol{v}^*$ is the tentative velocity. Eq. (25) is known as the predictor and calculates a tentative velocity by accounting for viscous and external forces. Eq. (26) is the pressure Poisson equation, which computes the pressure at the next time step from this tentative velocity. For single-phase flows with constant density, the pressure Poisson equation reduces to the standard Laplacian form $1/\rho_0 \nabla^2 p$. Finally, Eq. (27) is referred to as the corrector and updates the velocity field for the subsequent time step. In this study, Eqs. (25)-(27) are discretized using the LSSPH given by Eq. (13). Thus, ISPH can be regarded as a semi-implicit formulation in which the pressure calculation involves an implicit solution of an elliptic problem. Because pressure is obtained from the incompressibility constraint rather than propagated through an artificial equation of state, the acoustic time-step restriction associated with the numerical speed of sound is removed. The remaining CFL condition is therefore governed primarily by fluid velocity and may be written as follows:

$$\Delta t \leq C_{\mathrm{CFL}} \left(\frac{h}{\max\limits_i \|\boldsymbol{v}_i\|_2}\right) \tag{28}$$

By relaxing the CFL condition in this manner, ISPH can employ time-step sizes several times larger than those typically used in explicit WCSPH simulations [49]. This relaxation comes at the cost of solving a global pressure Poisson equation at every time step. The resulting sparse linear system is generally solved using an iterative linear solver, increasing the computational cost per time step relative to a fully explicit update. Nevertheless, the larger allowable time-step size can compensate for this additional cost [43], and the total computational efficiency therefore depends on the problem, spatial resolution, and linear-solver performance.

## 3. SOFTWARE ARCHITECTURE AND SOLVER COMPOSITION

In CraftSPH, users construct problem-specific SPH solvers by selecting and combining numerical components rather than following a fixed solver procedure. To help users assemble these procedures intuitively, particle data, numerical operations, and the user-defined update sequence are managed separately. This organization allows components to be reused across solver configurations and makes it easier to add or modify a numerical method without changing unrelated parts of the solver. Section 3.1 describes the management of particle data and computational state, Section 3.2 explains how numerical components are connected, and Section 3.3 describes their assembly into a solver and the scope of AD. The numerical formulations of the individual methods are described in Section 2 and the cited references.

### 3.1. Particle data and computational state

SPH solvers combine particle types and physical or auxiliary variables that differ across problems. Passing each quantity separately to every numerical operation would couple the solver procedure to a particular data layout and make later extensions harder to maintain. CraftSPH instead separates analysis settings and particle data from the numerical operations: configuration files specify the case, while `SPH_Particle` holds the shared computational state passed to the user-defined solver. This permits components to use the same state-management mechanism when new particle variables or computational procedures are introduced.

#### 3.1.1. Variables and computational state

An instance of `SPH_Particle` is denoted below by particle. Users can define physical and auxiliary variables according to the target problem by specifying their names, tensor ranks, and data types. Listing 1 shows an example of variable definitions. A registered velocity variable `u` can then be accessed as `particle.u`. This enables users to add the particle-wise quantities required by a new physical model or numerical method and make them available to multiple numerical components through the same interface, without rebuilding the solver's particle-data structure.

The particle object also stores the number of particles, physical time, step number, time-step size, and information used in neighbor search. Components can therefore refer to a common computational state rather than maintain duplicate copies of these values. Output selection follows the same registered variables and particle types and remains separate from the numerical update sequence.

**Listing 1:** Examples to define variables

```
p: # pressure
  order: 0
```

```
  dtype: float
u: # velocity
  order: 1
  dtype: float
rho: # density
  order: 0
  dtype: float
m: # mass
  order: 0
  dtype: float
```

### 3.1.2. Particle types and sets

CraftSPH separates the definition of particle variables from their storage across particle types. Users can define the particle types (objects) required by a problem and combine them into named particle sets (object_sets), rather than being restricted to a fixed selection. A velocity variable can be accessed for fluid and wall particles as `particle.u.fluid` and `particle.u.wall`, or for a user-defined set of both types as `particle.u.all`. Listing 2 illustrates these particle-type and set definitions. The particle object handles concatenation when a set is read and partitioning when values are assigned to that set, so numerical components can specify the participating particle sets without implementing these tensor operations themselves. The same mechanism applies when particle types are added or removed at open boundaries, with registered variables updated consistently.

With this management strategy, the solver does not need to manage tensors for each particle type directly. The addition of particle types and changes in their combinations are mainly handled through the configuration file, and users only need to specify which particle types are used as computational targets. This allows the computational procedure described below to focus on the physical model and numerical method.

The configuration identifies the analysis settings and user-defined solver to be used for a case. At initialization, an `SPH_Particle` object is created and passed to `Solver`, and subsequent calls to `Solver.step()` advance the shared state. The detailed input syntax and execution arguments can be checked in the public implementation.

**Listing 2:** An example to define `objects` and `object_sets` in the configuration file.

```
objects:
  - id: 0
```

```yaml
    key: "fluid"
    path: "./path/for/fluid.h5"
  - id: 1
    key: "wall"
    path: "./path/for/wall.h5"
  - id: 2
    key: "VM"
    path: "./path/for/VM.h5"
object_sets:
  all: ["fluid", "wall"]
```

### 3.2. Composable numerical components

CraftSPH allows users to define the particle objects and field variables required for their analyses and to select numerical operations according to the target problem. The numerical components use the shared particle state and neighbor information without fixing a particular combination of particle objects, fields, or operations within the solver. This structure enables users to construct problem-specific solvers by combining the required components. When more advanced or problem-specific schemes are required, users can implement them through the corresponding component interfaces and incorporate them into their own solvers. Supporting such user-driven extensions is an important design objective of CraftSPH.

#### 3.2.1. Shared particle interactions and differential operators

A background-grid bucket search constructs neighbor relationships between particle sets selected by the user. The source and target sets are specified separately, and the resulting neighbor information is stored as an object named in the form `source2target`. For example, a search with `fluid` particles as source and `wall` particles as the target is represented by `fluid2wall`. Each neighbor object stores the selected indices and the associated relative positions and distances. Kernel functions and differential operators use this shared interaction data. Thus, neighbor-index generation is separated from the continuous geometric and numerical evaluations performed on selected pairs. `KernelFunction` selects an implemented kernel, while a newly implemented kernel value and derivative can be registered through the same selection interface. Replacing or extending a kernel need not change the particle-state representation or the solver-side sequence, although the new implementation and its numerical suitability must still be checked.

The classical SPH, CSPH, and LSSPH approximations described in Section 2 are provided through differential-operator interfaces. In the implemented formulations, corrections to kernel and differential weights allow the solver to request a physical derivative without reproducing each approximation procedure in its update sequence. This keeps the choice of spatial discretization distinct from the governing terms assembled by the solver and makes it possible to maintain an existing operator or add a compatible new one locally. In addition, CraftSPH supports matrix form of these major differential operators, which are required for implicit schemes. In the matrix form, coefficient components among source particles and components originating from `boundary` particles are constructed separately. Here, boundary particles refer to particles within the set of target particles, in the context of the `source2target` neighbor relationship, that are not source particles. By storing the boundary components separately, Dirichlet and Neumann conditions can be imposed later from the boundary condition module. A new differential model can be added by implementing and registering the required operator interface, while retaining the existing calling pattern in the solver. Its accuracy and compatibility with the selected numerical components must then be evaluated for the intended problem.

### 3.2.2. Main computations, stabilization techniques, and boundary conditions

To support the construction of problem-specific solvers, CraftSPH provides the differential operators described above, stabilization techniques, and boundary treatments as separate components rather than fixing their combination in a single solver. Table 1 summarizes the components provided with CraftSPH as a starting point for such solver construction. These components use the shared particle state and neighbor information, whereas their selection and order in the computational procedure are specified by the user-defined solver described in Section 3.3. The examples in Sections 4 and 5 demonstrate selected configurations, rather than every component listed in Table 1 or every possible combination. Detailed arguments and return values for these components should be checked against the public implementation corresponding to the reported results.

Stabilization techniques also use the particle and neighbor data already available to the solver. Keeping corrections separate from the update sequence allows their implementation to be maintained or replaced without embedding a fixed stabilization choice in every solver. The user selects where to apply each compatible correction and evaluates its effect on the stability and accuracy of the resulting solver for the problem at hand.

Boundary treatments are connected both to particle-state management and to numerical operators. Wall values may be supplied through ghost-particle interpolation [50], whereas inlet and outlet operations add, reclassify, or remove particles together with their registered variables. This keeps the state used by subsequent SPH operations consistent as particle sets change, rather than requiring each solver component to reconcile changed tensor lengths separately. These boundary condition treatments support two types of use. Firstly, physical quantities are directly interpolated or corrected. Secondly, boundary conditions are incorporated into the coefficient matrix or right-hand side vector. The former is used to evaluate velocity and pressure values of wall particles. The latter is used when solving linear systems in an implicit scheme. This structure allows the same boundary condition treatments to be reused not only in explicit computation but also in implicit computation. When a calculation requires particle classification, Boolean masks can be used to select the particle groups to which boundary treatments or other numerical operations are applied. For example, the implemented multistage detection [40] provides particle classifications for free-surface flows. The classification and mask-based selection are discrete and should be distinguished from the differentiation of continuous numerical operations on the selected particles.

Table 1. Numerical components available in CraftSPH and their role in solver composition.

| Contents | Name | Class |
|---|---|---|
| Kernel functions [2][51][52] | Gaussian<br>Cubic spline<br>Quintic spline<br>Wendland C2<br>Wendland C4<br>Spiky<br>Poly6 | `KernelFunctions` |
| Differentiable operators | Classical SPH [1][2]<br>CSPH [33][34]<br>LSSPH [5] | `FunctionAndOperators` |
| Time integrations | Euler<br>Runge-Kutta<br>Adams-Bashforth | `TimeIntegration` |
| Particle shifting | Dynamic stabilization (DS) [53]<br>Advective dynamics stabilization (ADS) [54]<br>Optimized particle shifting (OPS) [7]<br>Density-based particle shifting (DPS) [55]<br>Surface fitting [41]<br>Dynamic pair-wise particle collision (DPC) [56] | `FunctionAndOperators` |
| Stabilization | XSPH [46]<br>Artificial viscosity [44]<br>Density diffusion [6]<br>Vortex viscosity [57] | `FunctionAndOperators` |
| Linear solver | CG [58]<br>BiCGStab [59]<br>BiCGSafe [60] | `FunctionAndOperators` |
| Pressure treatment | Tait equation<br>Symmetric pressure force [61] | `FunctionAndOperators` |

### 3.3. Solver construction and support for automatic differentiation

Users construct a problem-specific Solver by arranging numerical components in `step()` around the shared particle state. Listing 3 illustrates neighbor search, operator evaluation, and state updates in a simplified procedure. The governing terms, boundary treatments, and stabilization techniques can be selected according to the problem while reusing the same particle-state and differential-operator interfaces. This approach applies to both explicit procedures based on direct operator evaluations and implicit procedures involving matrix assembly and linear solves. `TimeIntegration` supplies state-update formulas, while the user-defined solver arranges the required evaluations, including any stage-wise neighbor search and kernel re-evaluation, in a numerically consistent sequence. In the fluid-flow examples of Sections 4 and 5, WCSPH evaluates pressure using an equation of state, whereas ISPH constructs and solves a pressure system with boundary contributions before applying the pressure correction. Because the numerical operations and update sequence are separated, individual components can be maintained or replaced without rewriting the entire procedure.

Numerical operations performed on PyTorch tensors can contribute gradients to an inverse-analysis loss or connect an SPH computation with a learning model when a differentiable path exists. CraftSPH uses PyTorch autograd, and users can control which parts of a solver procedure are recorded for differentiation using familiar operations such as `torch.enable_grad()` and `torch.no_grad()`. The entire forward calculation therefore need not be recorded, provided that the dependencies required for the derivative of interest remain connected. For example, neighbor indices may be selected without differentiating the selection itself, while numerical evaluations on the selected particle pairs remain in the computational graph. However, excluding a dependence that affects the result can change the gradient obtained by AD. Section 5 identifies the paths used in the inverse analyses and discusses the consequences of the choices made there, including the treatment of the ISPH pressure solve. The component structure permits a learning model to exchange tensor-valued quantities with the SPH solver, but this study does not report a new neural-network experiment.

**Listing 3:** Simplified pseudocode showing the computational procedure from neighbor search and force evaluation to particle position update. `self.particle` denotes the particle object created during initialization and passed to solver. The objects and variables used in this listing are defined in Listing 1 and Listing 2.

```
# neighbor search
```

```
NeighborSearch.update_neighbor(self.particle, source="fluid",
target="all")

# prepare class instance
self.func_op = FunctionsAndOperators(self.particle, x2y="fluid2all",
density="rho", mass="m")

# calculate force
f_p = -self.func_op.gradient("p")/self.particle.rho.fluid
f_vis = self.func_op.artificial_viscosity("u")

# velocity and position update
self.particle.u.fluid = TimeIntegration.Euler(self.particle.u.fluid, (f_p
+ f_vis), self.particle.dt)
self.particle.xyz.fluid = TimeIntegration.Euler(self.particle.xyz.fluid,
self.particle.u.fluid, self.particle.dt)
```

## 4. EXPERIMENT FOR FORWARD ANALYSIS

This section evaluates the basic analytical performance of forward simulations and the effectiveness of the implemented numerical modules. In these tests, the ISPH scheme is used with LSSPH for spatial discretization. This shows that CraftSPH is not limited to standard explicit schemes or classical SPH formulations but can build advanced solvers by integrating multiple numerical modules. The validation problems considered are Poiseuille flow, rising-bubble, and two and three-dimensional dam-break. These problems represent, respectively, internal flow with inlet and outlet boundaries, multiphase flow, free-surface flow with large deformation, and three-dimensional flow with complex fluid and boundary geometries. Comparisons with analytical solutions, experimental data, and established benchmark results are used to evaluate the effectiveness of the accuracy and its applicability to a wide range of fluid-flow problems. Case-specific material properties and stabilization methods are summarized in the tables provided in the respective subsections. For additional details on the stabilization parameters, readers are referred to the corresponding case solvers in the publicly available CraftSPH v0.1.0 release.

### 4.1. Poiseuille flow

First, Poiseuille flow is considered to evaluate the accuracy of internal-flow simulations and the validity of the inlet and outlet boundary treatments. Poiseuille flow is a fully developed laminar flow between two parallel plates, and its steady-state velocity profile can be obtained analytically. This makes it a representative benchmark for evaluating numerical accuracy under internal-flow conditions. The analytical solution for the streamwise velocity of steady Poiseuille flow between parallel plates is expressed as follows:

$$u(y) = 4v_{\max}\frac{y}{H}\left(1-\frac{y}{H}\right) \tag{29}$$

where $H$ is the distance between the parallel plates and $v_{\max}$ is the maximum velocity, set to 1 $\mathrm{m/s}$ in this test. The computational domain had a streamwise length of 1.0 m and a channel height of $H = 0.2$ m. No-slip boundary conditions were imposed on the upper and lower walls. Twenty sensors were uniformly placed along the channel height at $x = 0.5$ to compare the computed velocity profile with the analytical solution. Other simulation parameters and selected numerical methods are summarized in Table 2.

Table 2. Parameters and numerical methods used for the Poiseuille-flow.

| Parameters | Value |
|---|---|
| Time step size [s] | 0.001 |
| End time [s] | 10.0 |
| Initial particle spacing $h_p$ [m] | 0.005 |
| Smoothing length [m] | $1.1h_p$ |
| Kernel function | Quintic spline |
| Density [$\mathrm{kg/m^3}$] | 1000 |
| Kinematic viscosity [$\mathrm{m^2/s}$] | $1.0 \times 10^{-6}$ |
| Linear solver | BiCGSTAB |
| Particle shifting technique | OPS |

Fig. 1 shows a snapshot of the flow field at $t = 10.0$ s, and Fig. 2 compares the velocity profile at the center of the channel with the analytical solution. The computed velocity profile showed good agreement with the analytical parabolic profile, including in the vicinity of the walls. In addition, although particles were continuously added at the inlet and removed at the outlet throughout the simulation, no noticeable numerical instability associated with the variation in particle number was observed. These results confirm that the analytical Poiseuille velocity profile can be stably maintained in an internal flow with inlet and outlet boundaries.

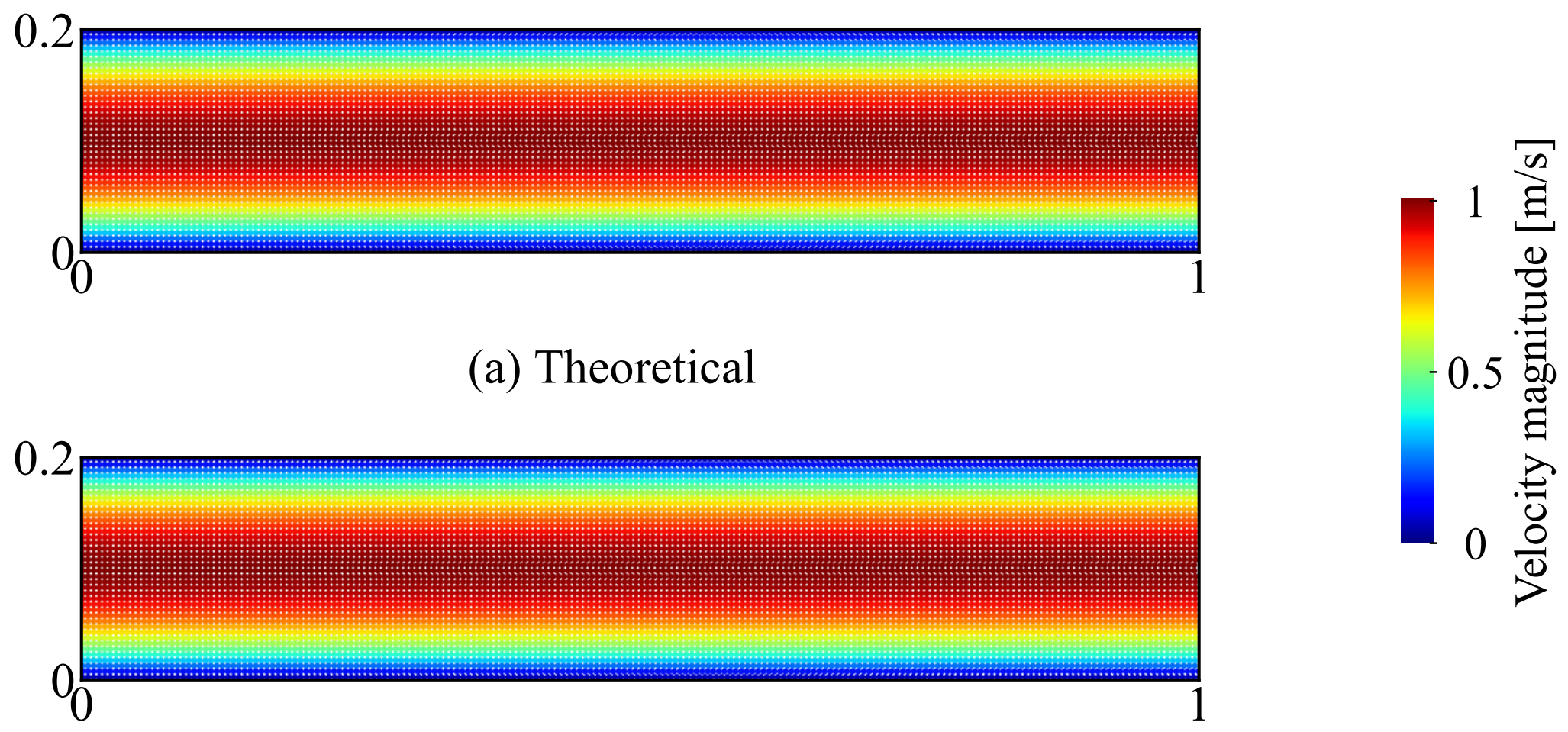

Fig. 1 Snapshots of the theoretical and predicted velocity fields at $t = 10.0$ s

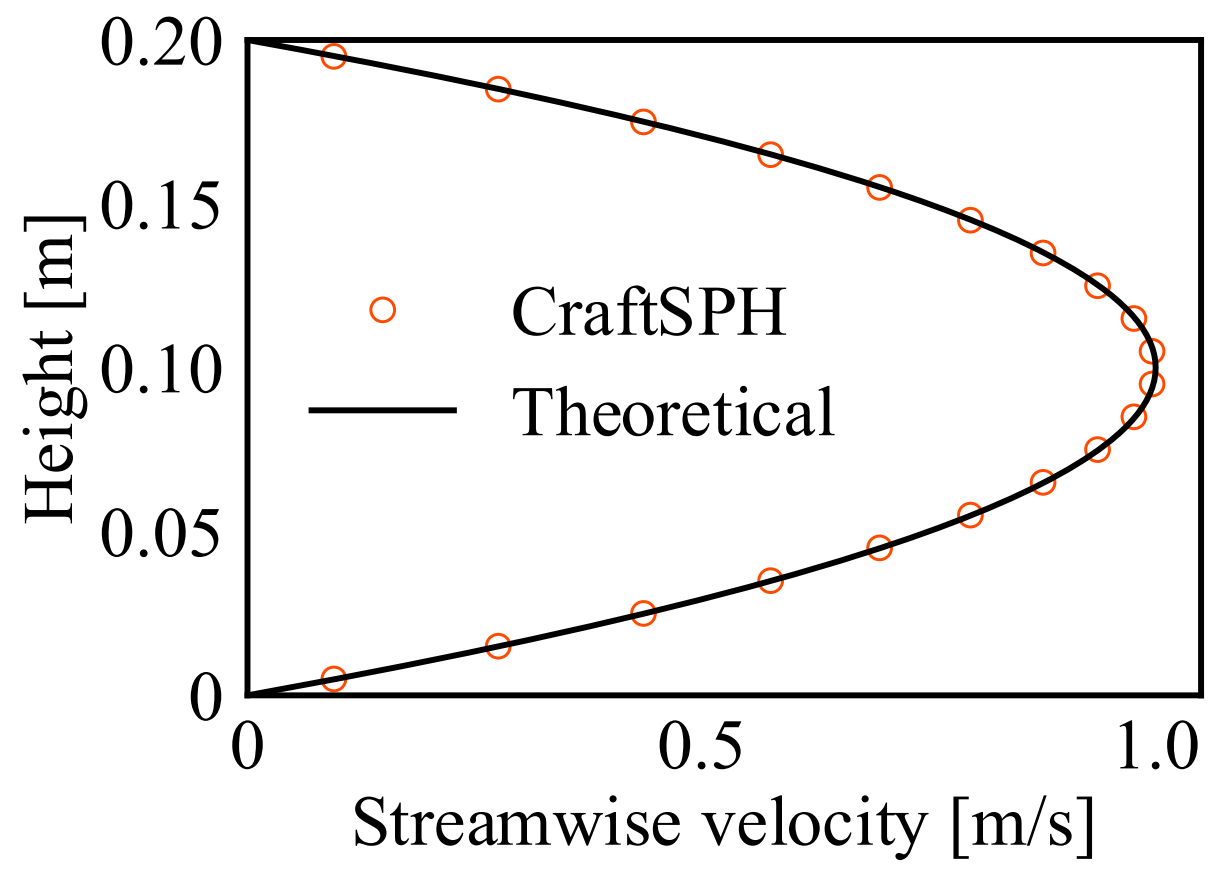

Fig. 2 Comparison of the theoretical and predicted streamwise velocity profiles at $x = 0.5$.

### 4.2. Two-dimensional dam break

The dam-break problem is a representative free-surface flow in which an initially stationary water column collapses under gravity and undergoes large deformation. The numerical setup follows the experimental conditions reported in a previous study [62]. Fig. 3 illustrates the computational domain, the initial water column, and the sensor locations. Other simulation parameters and selected numerical methods are summarized in Table 3.

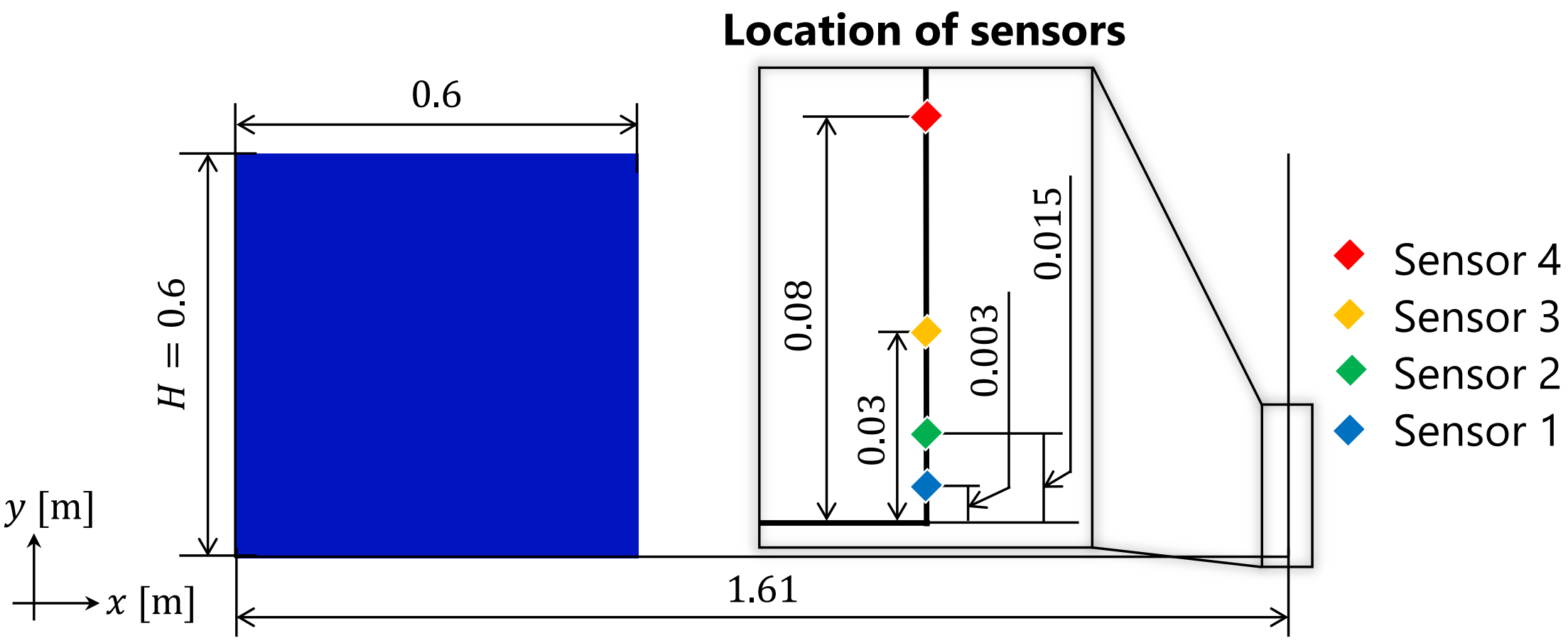

Fig. 3 Computational setup of the two-dimensional dam-break problem, including the computational domain, the initial water column, and the pressure sensor locations.

Table 3. Parameters and numerical methods used for the dam-break.

| Parameters | Value |
|---|---|
| Time step size [s] | $6.25 \times 10^{-5}$ |
| End time [s] | 2.0 |
| Initial particle spacing $h_p$ [m] | $1.25 \times 10^{-3}$ |
| Smoothing length [m] | $1.1h_p$ |
| Kernel function | Cubic spline |
| Density [kg/m$^3$] | 1000 |
| Kinematic viscosity [m$^2$/s] | $1.0 \times 10^{-6}$ |
| Gravity acceleration [m/s$^2$] | 9.81 |
| Linear solver | BiCGSTAB |
| Particle shifting technique | OPS / DPS / ADS |
| Stabilization technique | Vortex viscosity |
| Stabilization technique only for free surface and splash | XSPH / DPC |

Fig. 4 shows the flow field and pressure distribution at representative times, while Fig. 5 compares the pressure histories at the sensor locations with the experimental measurements. As shown in Fig. 4, after the initial water column collapsed, the fluid propagated along the bottom and formed a complex free surface with a large upward motion after impacting the downstream wall. Local negative-pressure regions were also observed within the fluid, particularly near $(x, y) \approx (1.6, 0.1)$ at $t = 0.8$ s and $1.0$ s, and near $(x, y) \approx (1.4, 0.1)$ at $t = 1.3$ s. These localized pressure reductions are qualitatively consistent with previous high-order particle-method simulations [40][41]. The pressure histories in Fig. 5 reproduce the rapid pressure increase caused by the fluid impact and the subsequent pressure fluctuations. Although some discrepancies are observed in the instantaneous peak values, the timing of the pressure rises, and the subsequent temporal variations are generally consistent with the experimental measurements.

As discussed in section 2.2.3, LSSPH can become unstable under strongly disordered particle distributions despite its high spatial accuracy. Nevertheless, CraftSPH allows a wide range of stabilization techniques to be readily incorporated, enabling stable simulations even for complex free-surface flows with strongly disordered particle distributions.

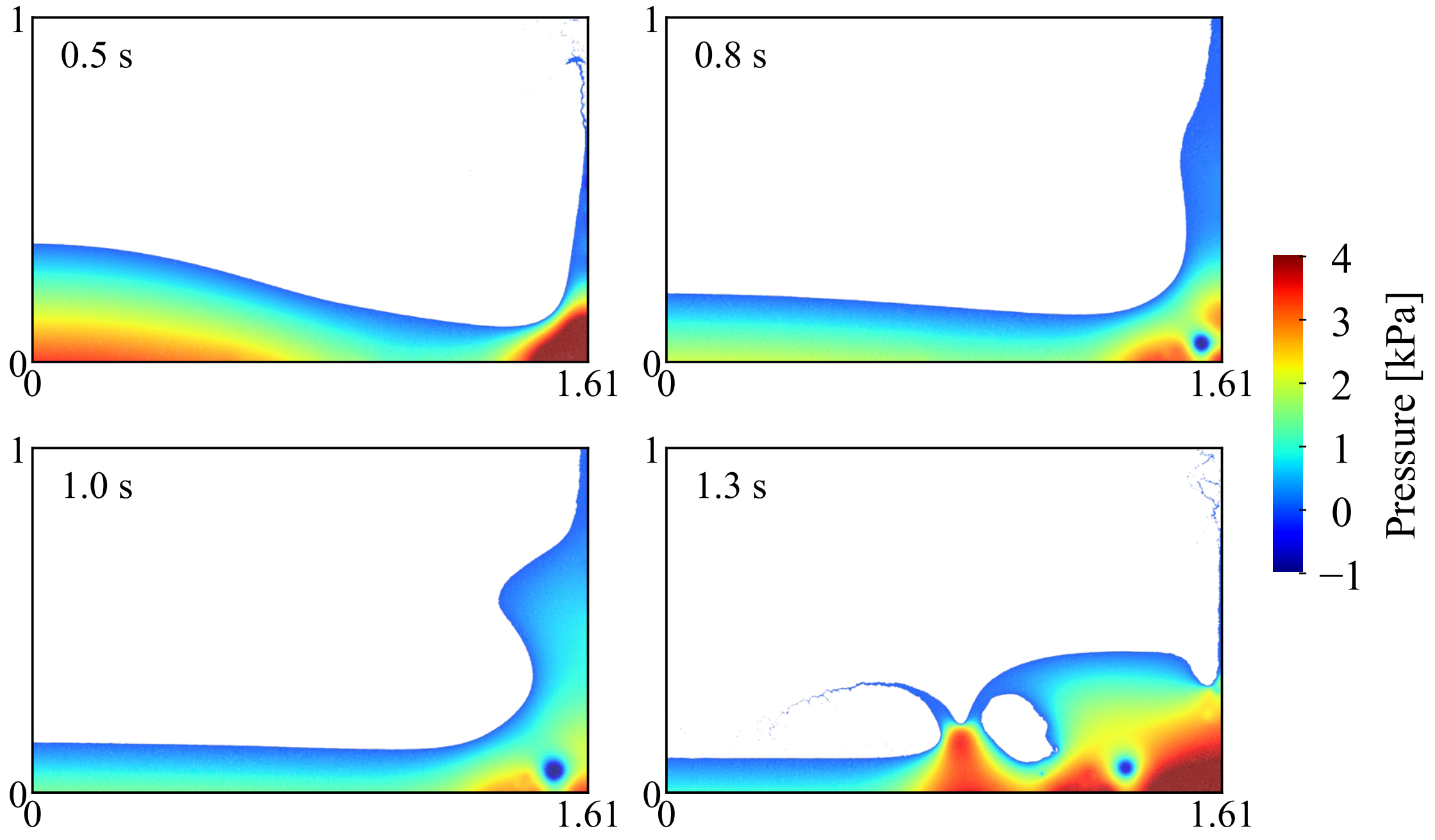

Fig. 4 Flow field and pressure distribution at representative times predicted by CraftSPH

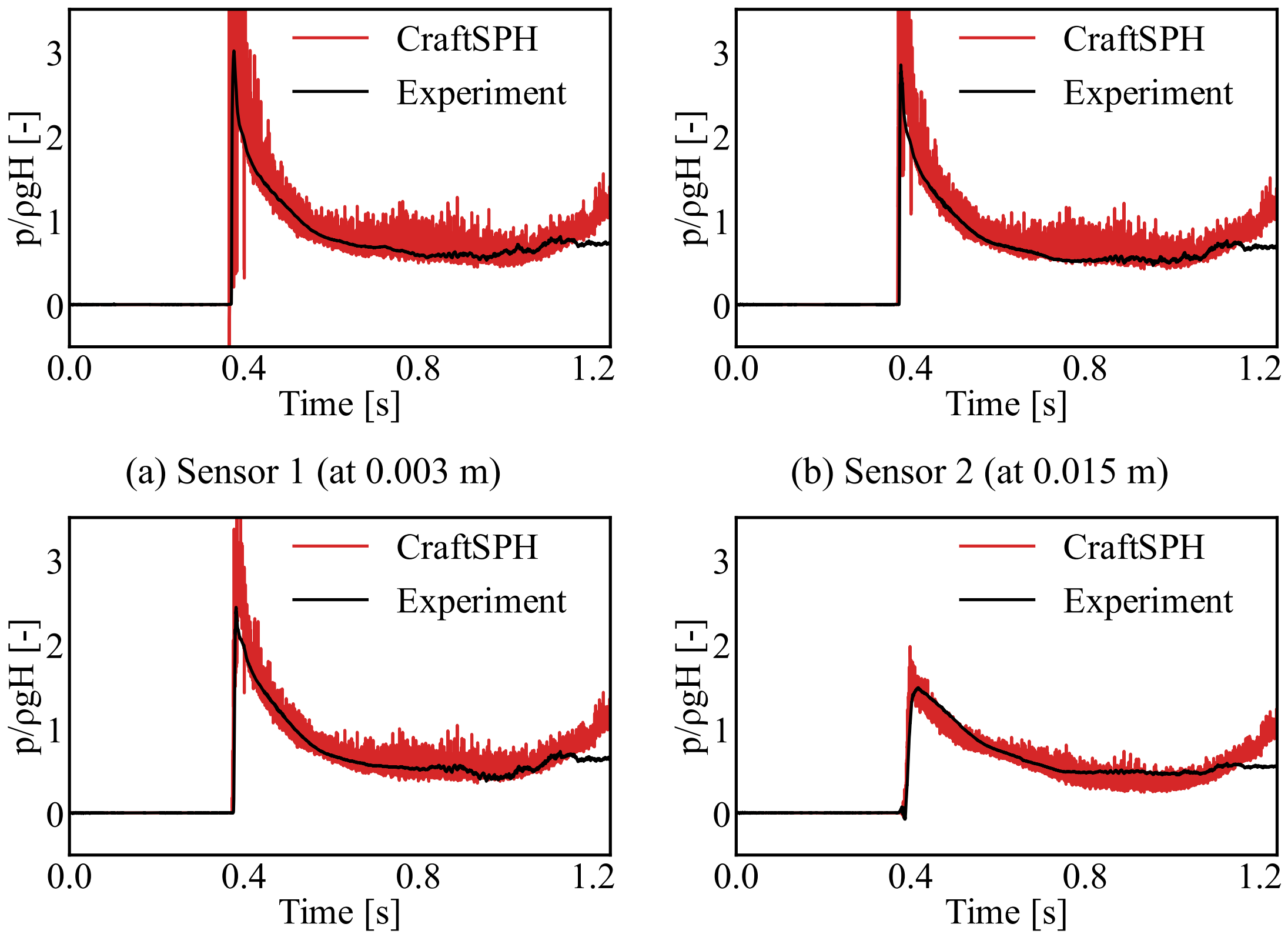

(a) Sensor 1 (at 0.003 m) (b) Sensor 2 (at 0.015 m)

(c) Sensor 3 (at 0.03 m) (d) Sensor 4 (at 0.08 m)

Fig. 5 Comparison of pressure-time history at each sensor between CraftSPH and experimental values.

### 4.3. Rising bubble

The rising-bubble is a representative multiphase-flow benchmark in which a bubble rises and deforms under buoyancy into two fluids with different densities and viscosities. The benchmark configuration proposed by Hysing et al. [63] is adopted. The benchmark includes Case 1 and Case 2, with Case 2 involving larger density and viscosity ratios and stronger interface deformation. Under such conditions, high-accuracy spatial discretization alone is insufficient, and specialized discretization and stabilization treatments are required. Therefore, to clearly evaluate the applicability of LSSPH discretization, Case 1 is adopted here.

The computational domain was a rectangle with a width of 1.0 m and a height of 2.0 m, as shown in Fig. 6 (a). A stationary circular bubble with a radius of 0.25 m was initially placed at $(x, y) = (0.5, 0.5)$. Surface tension was modeled using the continuum surface force model [64], with a surface-tension coefficient of 24.5 $\mathrm{N/m}$.

Table 4. Parameters and numerical methods used for a rising bubble.

| Parameters | Value |
|---|---|
| Time step size [s] | $5.0 \times 10^{-4}$ |
| End time [s] | 3.0 |
| Initial particle spacing $h_p$ [m] | 0.005 |
| Smoothing length [m] | $1.1h_p$ |
| Kernel function | Quintic spline |
| Density [$\mathrm{kg/m^3}$] (Bubble / Surrounding) | 100 / 1000 |
| Dynamic viscosity [Pa s](Bubble / Surrounding) | 1 / 10 |
| Linear solver | BiCGSAFE |
| Particle shifting technique | OPS |

Hysing et al. [63] reported three reference solutions that showed nearly identical results. Among them, the MooNMD solution was selected for comparison in the present study. MooNMD represents the interface using an ALE-based moving mesh, and its computational points move with the interface. This feature makes it closer to particle-based methods than the other two Eulerian approaches.

Fig. 6 shows (a) the bubble shape and pressure field at the final time and the time histories of (b) the center of mass, (c) circularity, and (d) rise velocity. The predicted bubble shape showed good agreement with the reference result, and a smooth pressure distribution was obtained as shown in (a). The time histories of the center of mass in (b) and rise velocity in (d) also agreed well with the reference solution, indicating that the rising motion of the bubble was accurately reproduced. However, a small discrepancy was observed in the circularity shown in (c). This is mainly due to the difficulty in evaluating the interface perimeter from discretely distributed particles. Since the bubble shape agrees well with the reference solution, the discrepancy is considered to be within the range of particle discretization errors. These results confirm that the ISPH-LSSPH solver constructed in CraftSPH can reproduce the bubble shape, displacement, and rise velocity in a multiphase flow consisting of two fluids with different densities and viscosities.

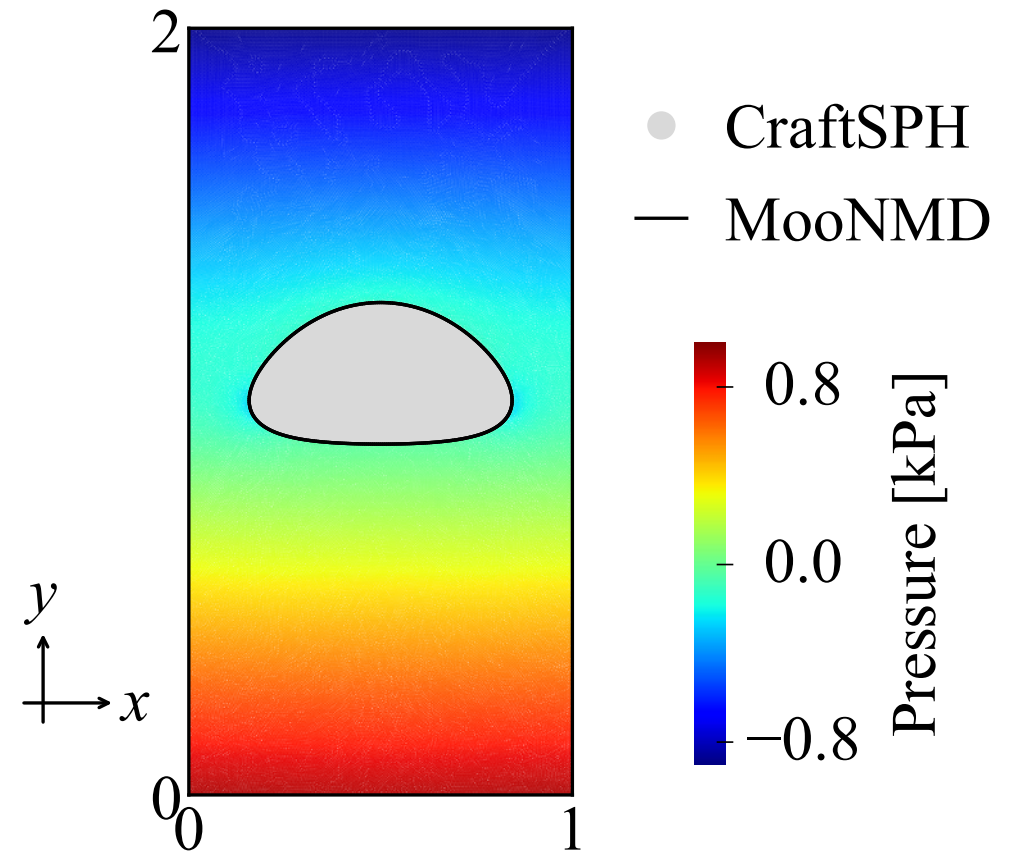

(a) Shape and pressure field at $t = 3$ s

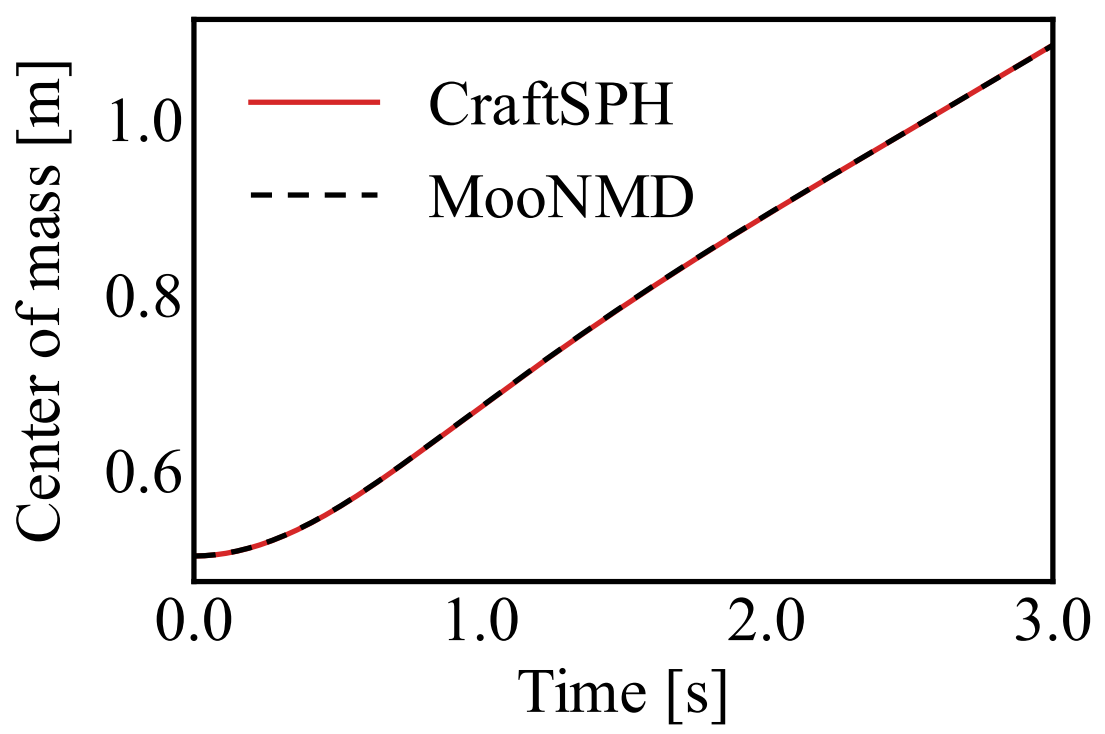

(b) Center of mass

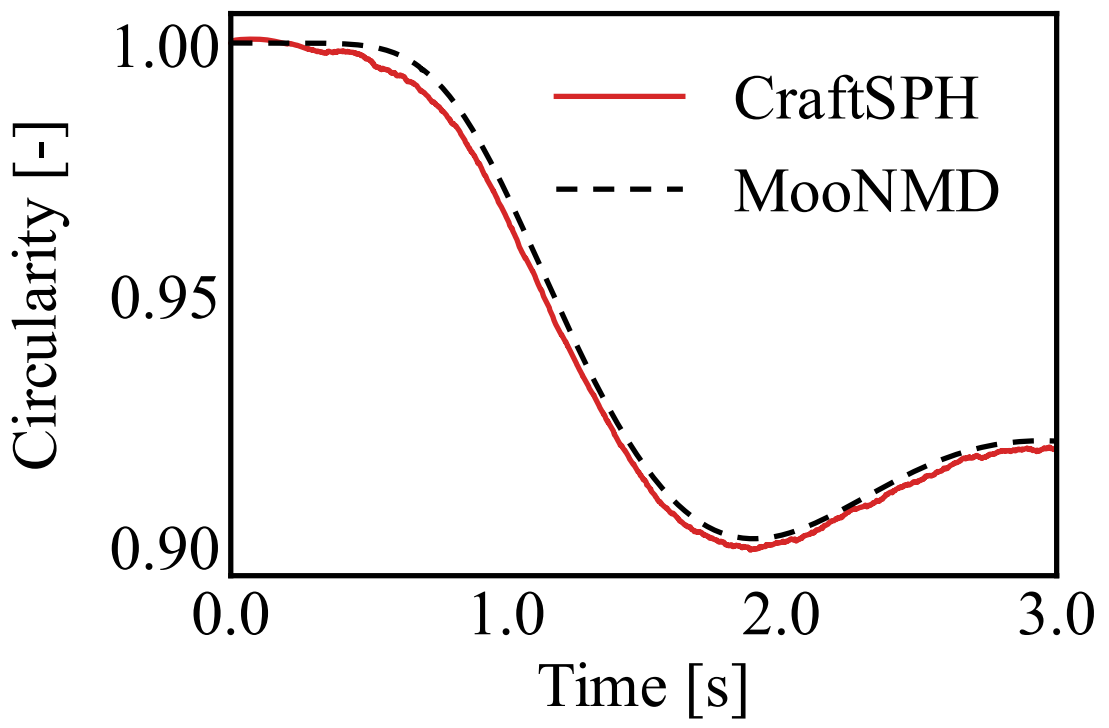

(c) Circularity

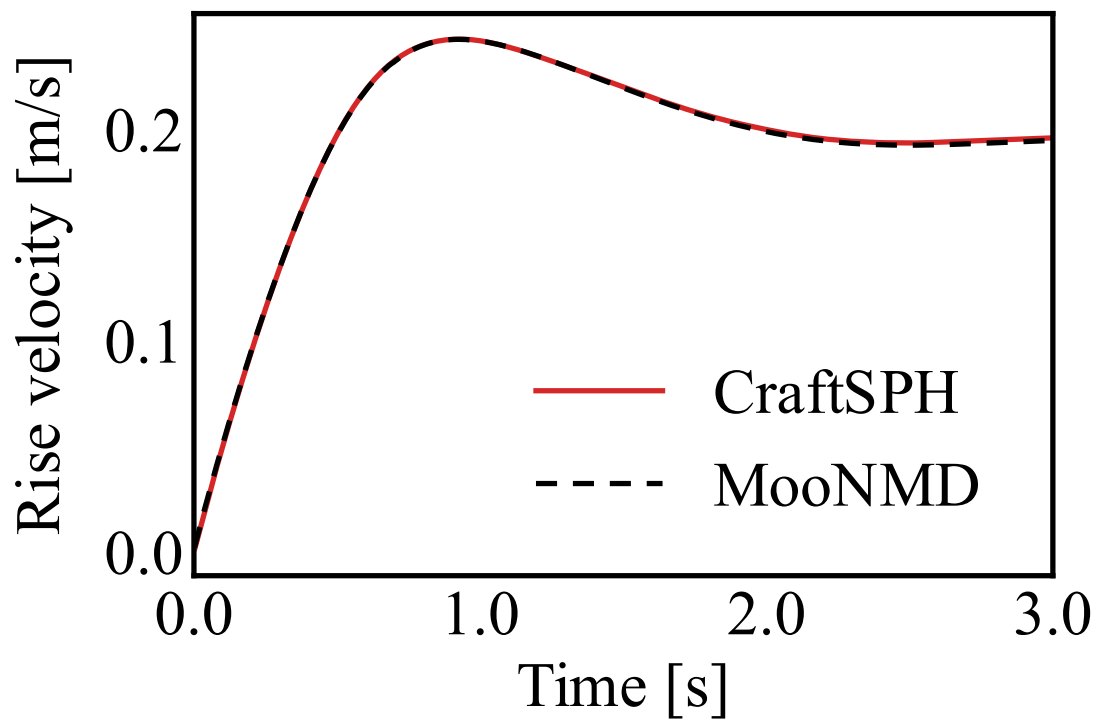

(d) Rise velocity

Fig. 6 Comparison of each quantitative metric between CraftSPH and reference solution in [63].

### 4.4. Three-dimensional dam break

Finally, a three-dimensional dam-break problem with complex fluid and boundary geometries is considered to evaluate the applicability of CraftSPH to three-dimensional flows. The present analysis was constructed based on the two-dimensional dam-break solver by modifying only the computational domain, boundary geometry, and initial conditions. All other major computational procedures were implemented using the same code as in the two-dimensional dam-break. The computational setup was based on the experiment of Kleefsman et al. [65]. The computational domain was a rectangular tank with dimensions of 3.22 m × 1.00 m × 1.00 m in length, width, and height, respectively. The initial water column was stationary and had dimensions of 1.228 m × 1.00 m × 0.55 m. A rectangular obstacle with a length of 0.16 m × 0.40 m × 0.16 m was placed on the bottom of the tank. The initial particle spacing was $2.0 \times 10^{-2}$ m. The time-step size was set to 0.001 s. Except for the initial particle spacing and time-step size specified above, the simulation parameters and numerical methods were the same as those listed for the two-dimensional dam-break problem in Table 3.

Fig. 7 shows the flow field and pressure distribution at representative times, and Fig. 8 compares the pressure time history at the sensor, located near the spanwise center of the upstream-facing vertical face of the obstacle and 0.025 m above the tank bottom, with experimental measurements. As shown in Fig. 7, the fluid propagated through the domain, impacted the obstacle, and flowed around it while forming a complex three-dimensional free surface. A spatially smooth pressure distribution was also obtained in the cross section, and the simulation remained stable throughout the three-dimensional analysis. As shown in Fig. 8, the computed pressure history reproduced the pressure rise associated with the arrival of the fluid and the subsequent pressure variations. Although discrepancy was observed in the onset time of the pressure rise, the subsequent pressure level and temporal variation were generally consistent with the experimental measurements. These results demonstrate the applicability of CraftSPH to complex three-dimensional free-surface flows. In addition, the solver developed for the two-dimensional dam-break problem was applied without any modifications to demonstrate the reusability of the modular solver design. If further improvement in accuracy is the primary objective, the boundary condition treatment, stabilization scheme, and their parameters may need to be individually selected and tuned for the present problem.

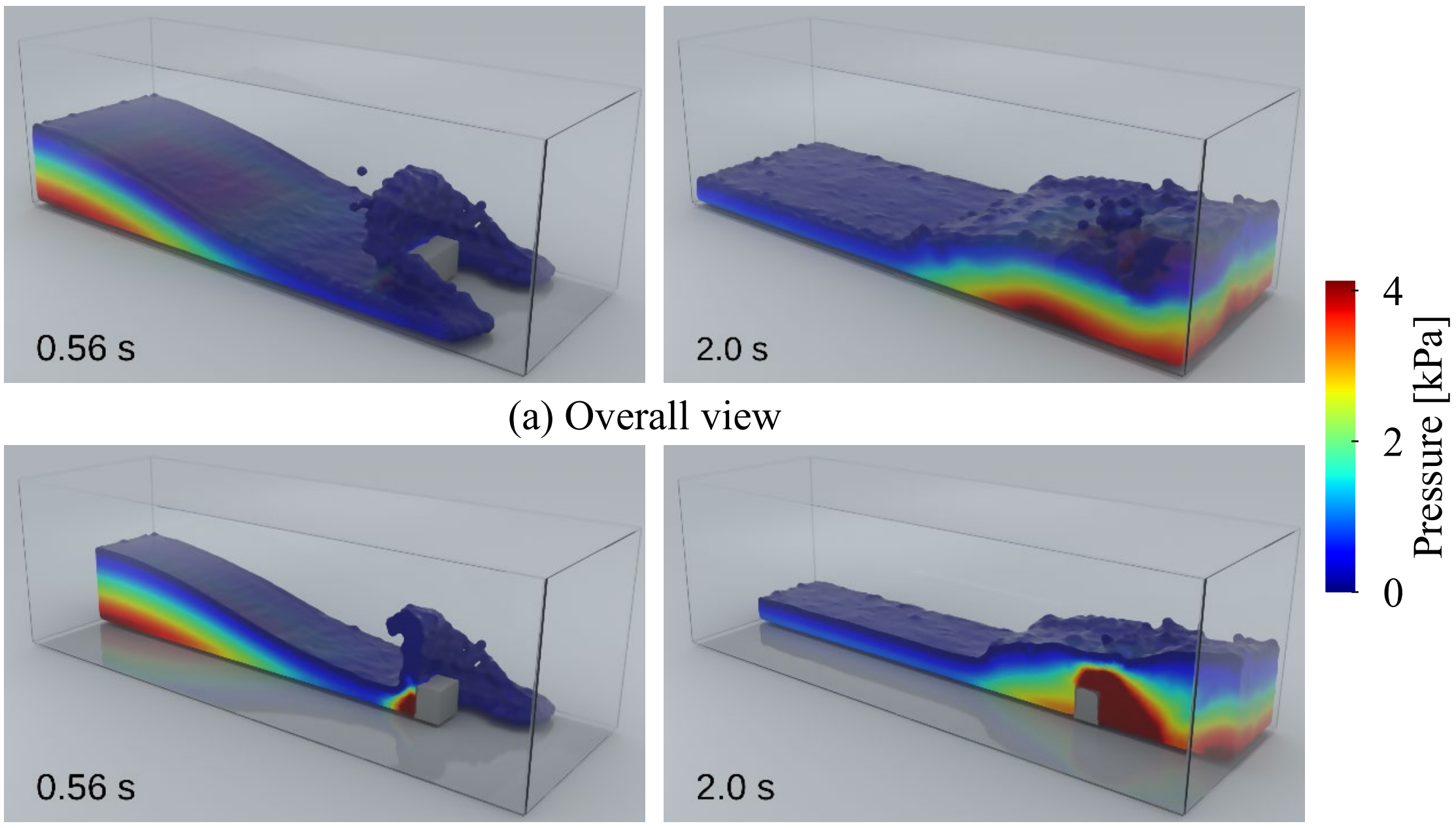

(a) Overall view

(b) Cross-sectional view

Fig. 7 Flow field and pressure distribution at representative times.

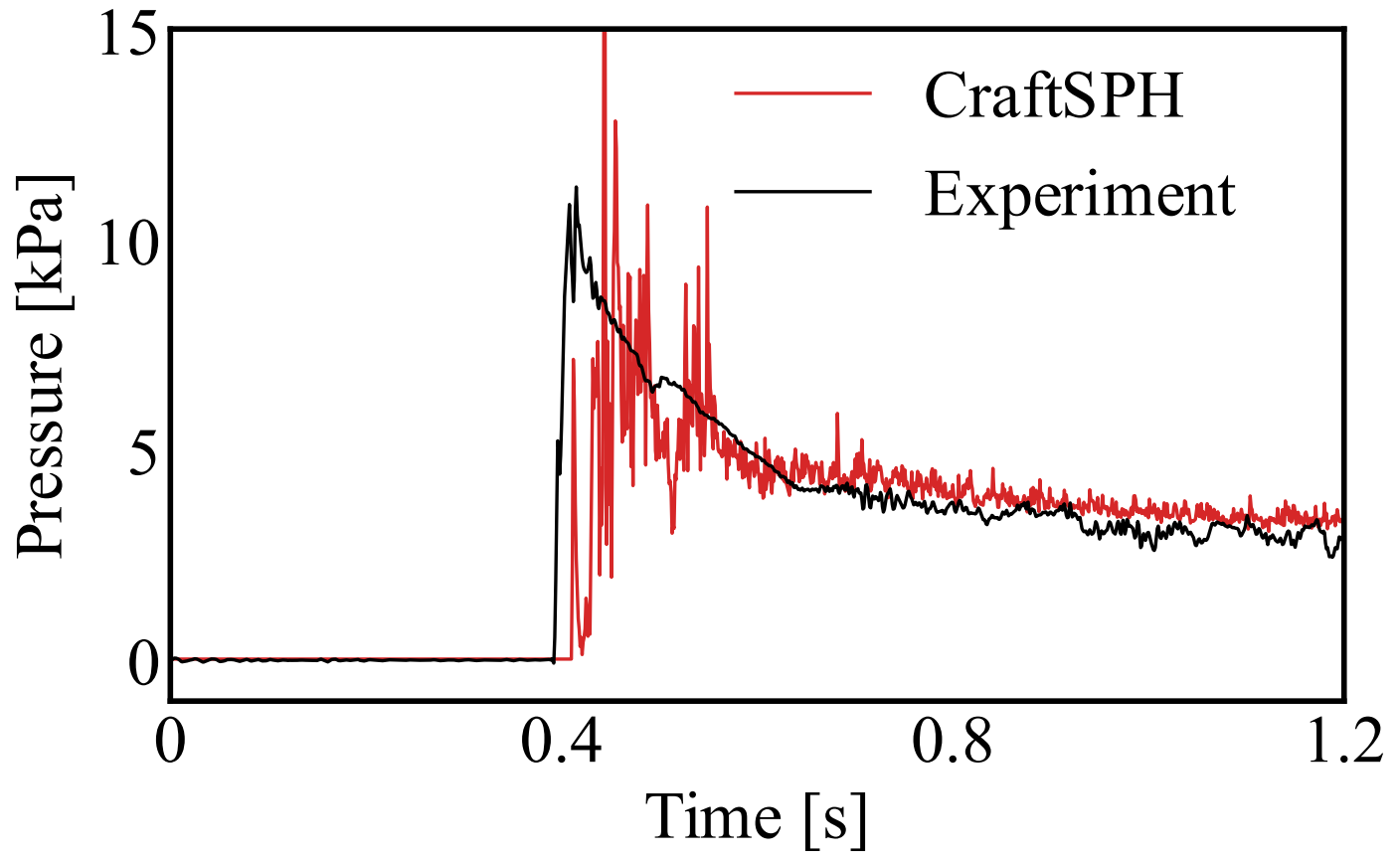

Fig. 8 Comparison of pressure-time history at sensor between CraftSPH and experimental values.

## 5. EXPERIMENT FOR INVERSE ANALYSIS

This section evaluates the effectiveness of AD in CraftSPH through inverse analysis. In the inverse analysis, physical parameters are estimated by minimizing the discrepancy between the observational data and numerical results, where the required gradients are computed using AD. Here, the kinematic viscosity is estimated for the Taylor–Green vortex, while the Reynolds number is estimated for the lid-driven cavity flow. Through these experiments, we evaluate the validity of gradient computation via AD in CraftSPH and its applicability to inverse analysis using reference data from established benchmark solutions. In addition, confirming that gradients can be propagated appropriately through the numerical modules provides a fundamental validation for integration with deep learning models.

The contribution of CraftSPH lies in its composable architecture, in which users construct problem-specific solvers from numerical components and define the scope of AD for each solver. Therefore, both WCSPH and ISPH are considered here using LSSPH to examine whether AD can be applied to both explicit and implicit SPH formulations with high-accuracy spatial discretization. However, making the entire solver differentiable is neither necessary nor always appropriate, because differentiating unnecessary procedures or repeated iterative solver operations may increase computational cost and cause gradient explosion or vanishing. Accordingly, the neighbor-search procedure is excluded from the computational graph for both schemes, and the iterative solution of the pressure Poisson equation is additionally excluded for ISPH. As in Section 4, readers are referred to the corresponding case solvers in the publicly available CraftSPH v0.1.0 release for stabilization parameters.

### 5.1. Estimation of kinematic viscosity in the Taylor-Green vortex

The Taylor-Green vortex is a representative benchmark problem in which vortical motion decays over time because of viscosity, and an analytical solution is available for incompressible flow. Since the decay of the velocity field strongly depends on the kinematic viscosity, the kinematic viscosity is estimated as an unknown parameter.

The Taylor–Green vortex has an analytical solution for the $x$-direction velocity $v_x$ and $y$-direction velocity $v_y$, given by follows:

$$v_x = V_0 \sin(kx) \cos(ky) \exp(-2\nu k^2 t) \tag{30}$$

$$v_y = -V_0 \cos(kx) \sin(ky) \exp(-2\nu k^2 t) \tag{31}$$

here, $V_0$ is the initial velocity amplitude and $k$ is the wavenumber. In the present study, $V_0 = 1$ and $k = \pi$ are used. Nine sensors are uniformly distributed over the computational domain. The kinematic viscosity is estimated by minimizing the discrepancy between the numerical solution interpolated at the sensor locations and the corresponding analytical solution. The computational domain and sensor locations are shown in Fig. 9. A slip condition is imposed on the boundaries based on Matsunaga et al. [66]. For each optimization step, the flow is first simulated from 0 to 1.0 s, and the kinematic viscosity is updated using the numerical solution obtained at 1.0 s. After the update, the simulation is restarted from 0 s and repeated until the prescribed number of epochs is reached. Other parameters for the simulations and inverse analyses are as shown in Table 5.

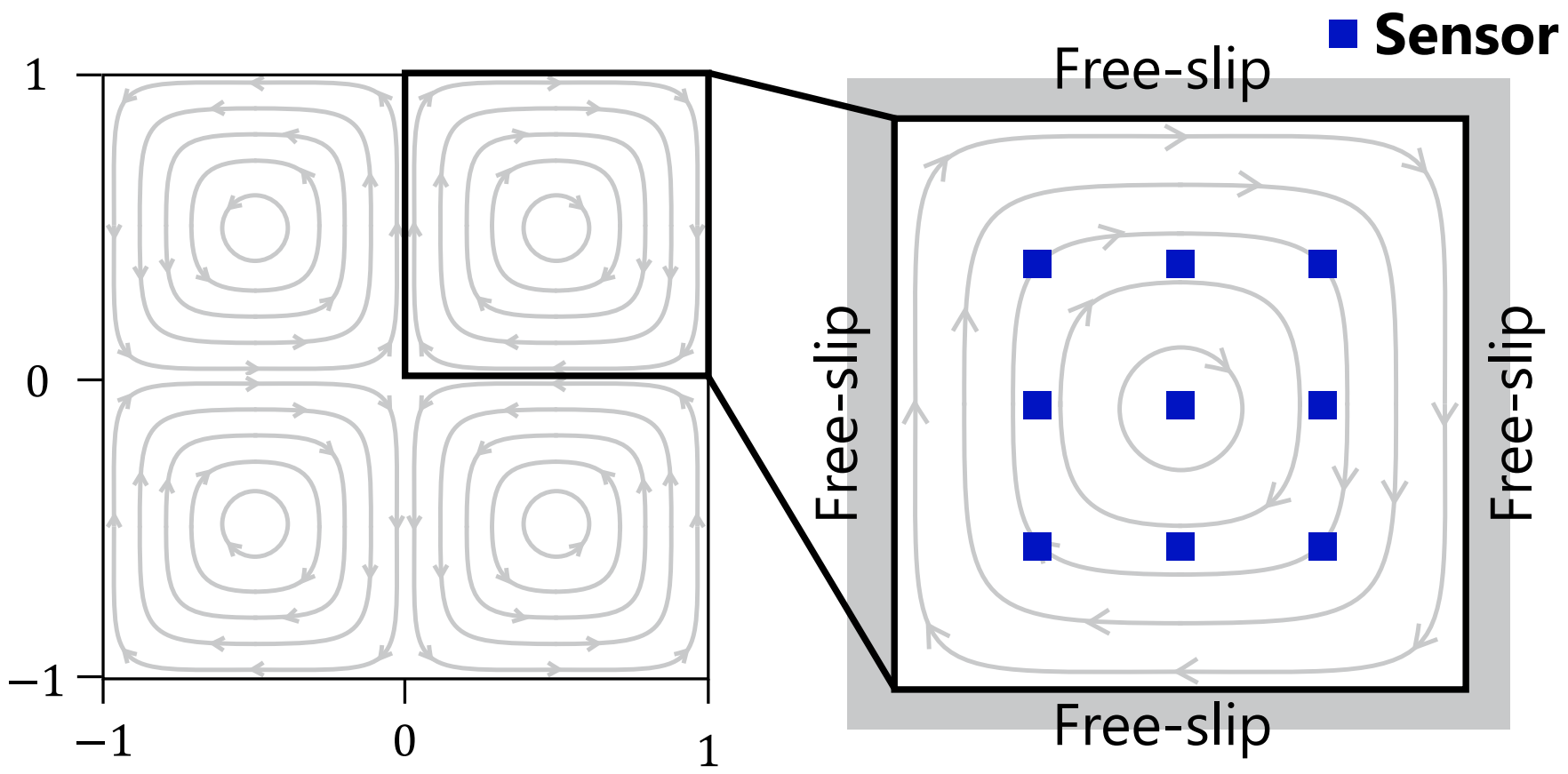

Fig. 9 Illustration of Taylor-Green vortex

Table 5. Parameters and numerical methods used for an inverse analysis of Taylor-

Green.

| Parameters | Value |
|---|---|
| Courant number [-] | 0.3 |
| End time [s] | 1.0 |
| Initial particle spacing $h_p$ [m] | 0.02 |
| Smoothing length [m] | $1.1h_p$ |
| Kernel function | Quintic spline |
| Density [$\mathrm{kg/m^3}$] | 1.0 |
| Kinematic viscosity [$\mathrm{m^2/s}$] (Initial / Target) | 0.001 / 0.01 |
| Speed of sound [$\mathrm{m/s}$] | 10.0 |
| Particle shifting technique | OPS |
| Optimizer | Adam [67] |
| Learning rate [-] | 0.1 |
| Epoch [-] | 300 |

Table 6 presents the estimated kinematic viscosity at the final time and the computational cost for each numerical scheme. Fig. 10 shows the convergence history of the estimated value, and Fig. 11 shows the velocity field at the final time. Computational time corresponds to the simulation up to 1.0 s, and the computational cost per time step was obtained by dividing the total computational time by the total number of time steps. For both WCSPH and ISPH, the estimated values converged to constant values, and the velocity fields at the final time were generally consistent with the theoretical solution. In particular, WCSPH achieved an estimation error of 0.19%, confirming that the kinematic viscosity can be identified with high accuracy using gradients obtained by AD. In contrast, the estimation error for ISPH was 12.13%, indicating lower estimation accuracy than that obtained with WCSPH.

To examine the difference in inverse-analysis results between WCSPH and ISPH, gradients of the loss with respect to kinematic viscosity were evaluated at the initial time step using AD and central finite differences (FD) and compared in Table 7. An FD perturbation of $10^{-6}$ was used, and relative errors were calculated with the FD gradient as the reference. For ISPH, the effect of excluding the pressure solve from the AD computational graph was also investigated. Pressure was solved at positive and negative perturbations of the kinematic viscosity, and each resulting pressure field was substituted into a calculation at the baseline condition. The difference between the resulting losses was used to estimate the pressure-response contribution to the gradient, which was then added to the AD gradient (AD+FD). The WCSPH AD gradient was found to be nearly identical to its FD counterpart, indicating that the change in loss with kinematic viscosity was captured by AD. In contrast, a relative error of 0.066% was observed between the ISPH AD and FD gradients. In ISPH, the velocity obtained in the prediction step is corrected by pressure. Consequently, a change in kinematic viscosity affects not only the predicted velocity but also the corrected velocity and the loss through its effect on pressure. When the iterative calculations involved in ISPH are excluded from the computational graph, however, this pressure response cannot be accounted for by AD. The difference from the FD gradient was nearly eliminated when the FD estimate of the pressure response was added to the AD gradient. Thus, the difference observed for ISPH appears to be largely attributable to the missing differentiation path through the pressure solve. These results indicate that the resulting gradient is affected by which operations are included in the AD computational graph.

Regarding computational cost, the total computational time was 14.5 s for WCSPH and 7.7 s for ISPH. Although ISPH requires a higher computational cost per time step because of the solution of the pressure Poisson equation, its less restrictive time-step constraint substantially reduces the required number of time steps.

These results demonstrate that CraftSPH can apply AD to both explicit and implicit numerical schemes and enables the effects of different computational schemes to be examined within the same differentiable framework.

Table 6 Predicted kinematic viscosity of each scheme in CraftSPH (target is $\boldsymbol{\nu} = \mathbf{0.01}$)

| Scheme | Computation time | | Prediction | |
|---|---|---|---|---|
| | Total [s] | 1 step [s] | Kinematic viscosity [$m^2/s$] | Error [%] |
| WCSPH | 14.5 | 0.009 | $9.98 \times 10^{-3}$ | 0.19 |
| ISPH | 7.7 | 0.056 | $8.79 \times 10^{-3}$ | 12.13 |

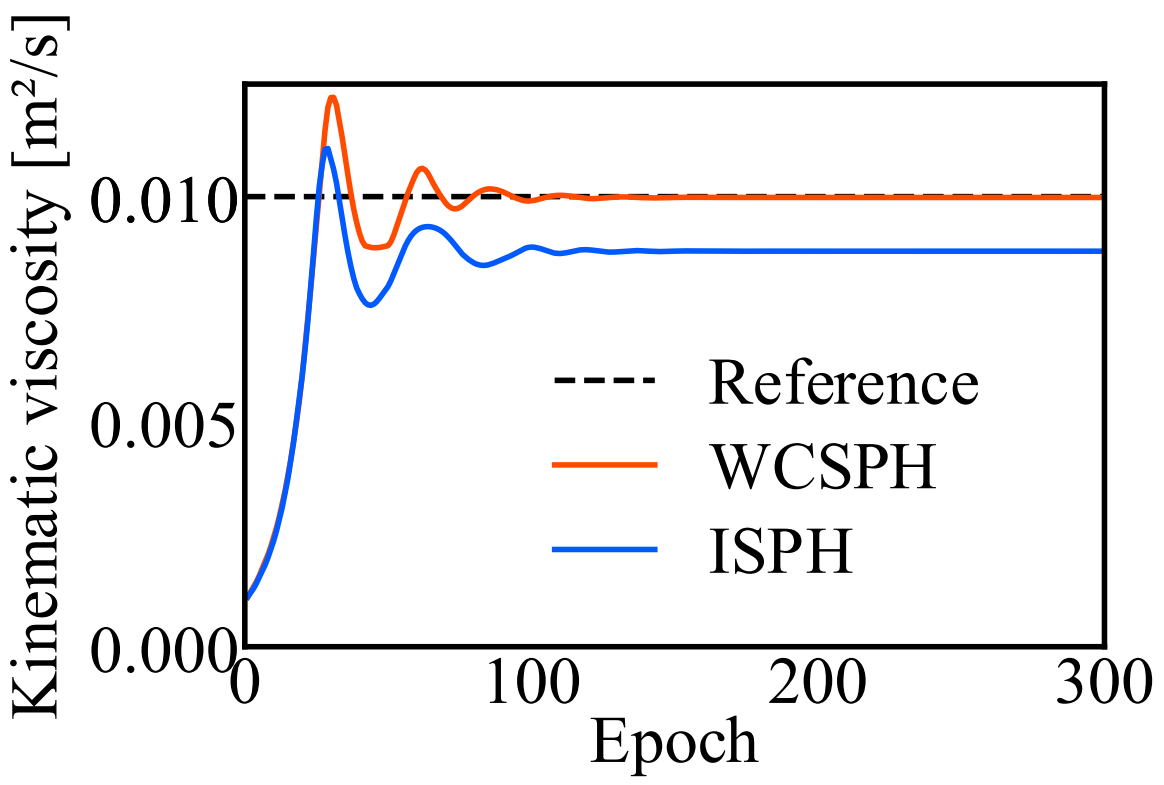

Fig. 10 Epoch history of predicted kinematic viscosity

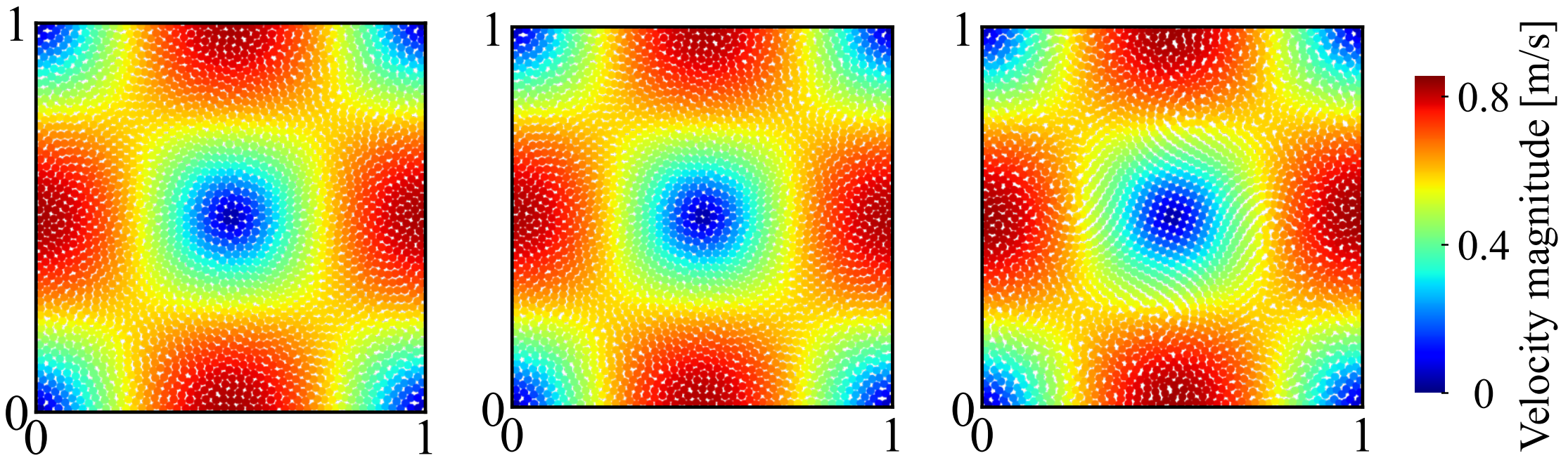

(a) Theoretical (b) WCSPH (c) ISPH

Fig. 11 Velocity field of each scheme at final time

Table 7 Comparison of loss gradients with respect to kinematic viscosity in the Taylor–Green vortex

| Scheme | FD | Prediction | Error [%] |
|---|---|---|---|
| WCSPH (AD) | $-1.5476 \times 10^{-4}$ | $-1.5476 \times 10^{-4}$ | $1.12 \times 10^{-8}$ |
| ISPH (AD) | $-1.4187 \times 10^{-4}$ | $-1.4178 \times 10^{-4}$ | $6.61 \times 10^{-2}$ |
| ISPH (AD+FD) | $-1.4187 \times 10^{-4}$ | $-1.4187 \times 10^{-4}$ | $1.28 \times 10^{-8}$ |

## 5.2. Estimation of the Reynolds number in lid-driven cavity flow

Lid-driven cavity flow is a representative benchmark problem in which the internal vortex structure and velocity distribution vary with the Reynolds number. In this experiment, a high-accuracy benchmark velocity distribution is used as observation data, and the Reynolds number is estimated to evaluate the applicability of CraftSPH to inverse analysis using reference flow-field data. The Reynolds number is defined as $Re = V_{\text{lid}}L/\nu$, where $V_{\text{lid}}$ is the lid velocity, and $L$ is the cavity length. In the present analysis, $L$ and $\nu$ are fixed, and $V_{\text{lid}}$ is treated as the optimization variable.

Ghia et al. [68] reported high-accuracy steady-state velocity profiles at several Reynolds numbers obtained from fine-grid numerical simulations. In the present study, the benchmark data at $Re = 3200$ are used as the observation data, and the Reynolds number is estimated from an initial value of $Re = 0$. Fig. 12 shows the computational domain and the sensor locations corresponding to the benchmark velocity data. Because the reference data represents a steady-state velocity field, the simulation is not restarted after each parameter update. Instead, Reynolds number is updated during the simulation until the prescribed end time is reached. Since changes in $V_{\text{lid}}$ require a finite time to propagate through the internal velocity field, the parameter is updated every 0.01 s for both WCSPH and ISPH. This ensures the same number of optimization steps over the same physical time, enabling a fair comparison between the two schemes. Other parameters for the simulations and inverse analyses are as shown in Table 8.

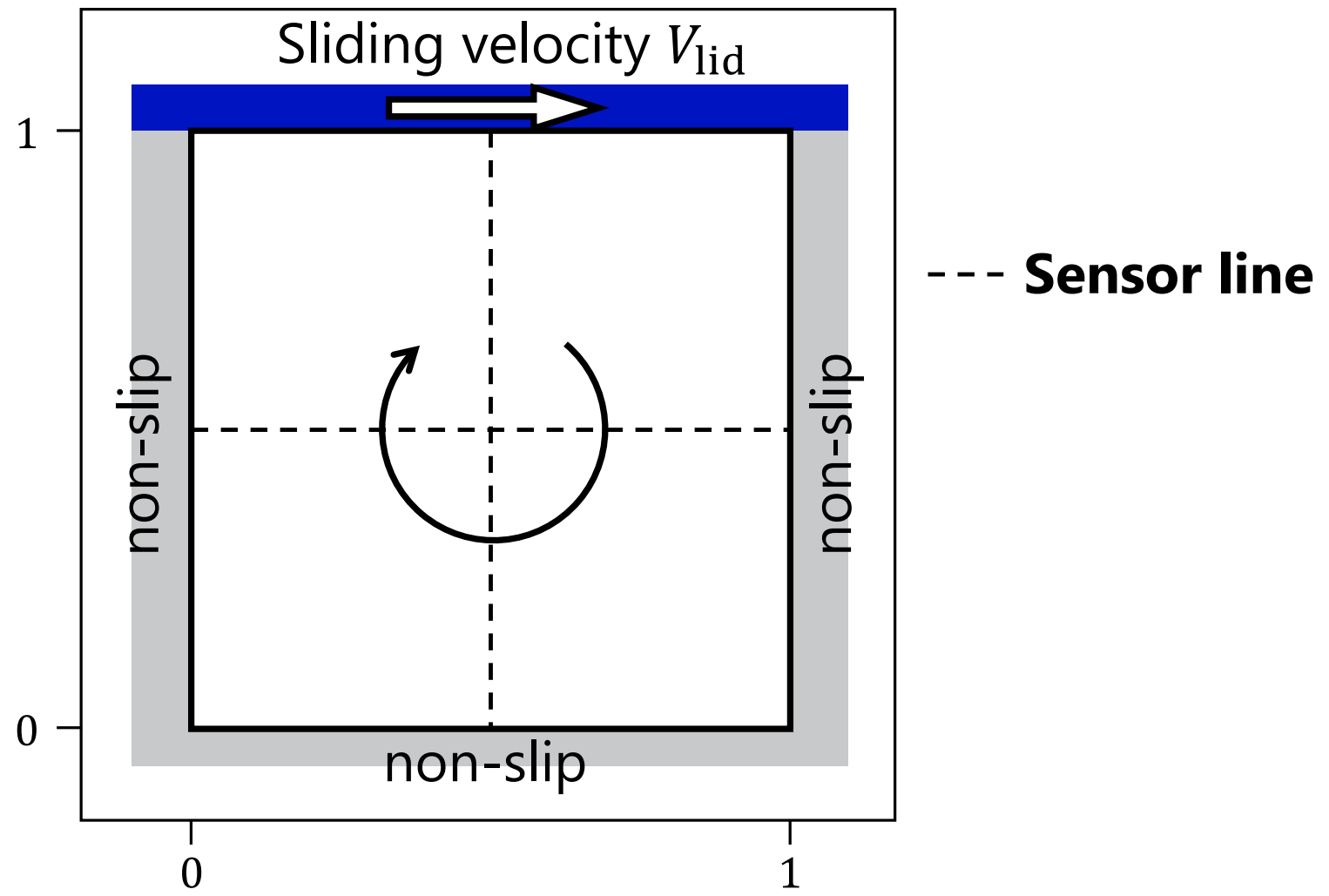

Fig. 12 Illustration of lid-driven cavity flow

Table 8. Parameters and numerical methods used for an inverse analysis of lid-driven Cavity flow.

| Parameters | Value |
|---|---|
| Courant number [-] | 0.3 |
| End time [s] | 30.0 |
| Initial particle spacing $h_p$ [m] | 0.01 |
| Smoothing length [m] | $1.1h_p$ |
| Kernel function | Quintic spline |
| Density $[\mathrm{kg/m^3}]$ | 1000 |
| Kinematic viscosity $[\mathrm{m^2/s}]$ | $1.0 \times 10^{-3}$ |
| Speed of sound $[\mathrm{m/s}]$ | 32.0 |
| Particle shifting technique | OPS / DPS |
| Optimizer | Adam |
| Learning rate [-] | 0.01 |

Table 9 presents the estimated Reynolds number at the final time and the computational cost for each numerical scheme. Fig. 13 shows the temporal history of the estimated Reynolds number, and Fig. 14 shows the velocity field at the final time. Although this inverse analysis is based on benchmark data obtained from an independent numerical method, the estimated Reynolds number converged well for both WCSPH and ISPH, and the final velocity fields exhibited generally similar distributions. The estimation errors were approximately 0.47% for WCSPH and 2.03% for ISPH. Although WCSPH still achieved higher estimation accuracy, ISPH also provided sufficiently accurate estimation, unlike the Taylor–Green vortex case.

To examine the difference in inverse-analysis results between WCSPH and ISPH, gradients of the loss with respect to lid velocity were evaluated at the first gradient-evaluation step using AD and central FD and compared in Table 10. An FD perturbation of $10^{-3}$ was used, and relative errors were calculated with the FD gradient as the reference. As in the preceding section, the pressure-response contribution for ISPH was also evaluated using FD and added to the AD gradient (AD+FD). As observed for the Taylor–Green flow, the WCSPH AD gradient was nearly identical to its FD counterpart. A relative error of approximately 0.009% was observed for the ISPH AD gradient, but the difference from the FD gradient was nearly eliminated when the FD pressure-response contribution was added. Despite this local gradient difference, the Reynolds number was estimated more accurately for the cavity flow than for the Taylor–Green flow. Thus, the accuracy of an inverse analysis cannot be judged solely from the local agreement between AD and FD gradients. When AD is used for inverse analysis, both the validity of the computed gradient and the accuracy of the resulting parameter estimate need to be assessed.

In contrast, ISPH allows a larger time-step size, resulting in a lower total computational time than WCSPH. These results indicate that high-accuracy parameter estimation can be achieved with an implicit computational scheme while maintaining computational efficiency, provided that the optimization target and the computational processes included in AD are appropriately selected. They also demonstrate that CraftSPH is applicable not only to idealized inverse problems based on analytical solutions but also to inverse analyses using established benchmark data obtained independently of the present solver.

Table 9 Predicted Re of each scheme in CraftSPH (target is $\boldsymbol{Re} = \mathbf{3200}$)

| Scheme | Computation time | | Prediction | |
|---|---|---|---|---|
| | Total [s] | 1 step [s] | Re [-] | Error [%] |
| WCSPH | 6,568 | 0.019 | 3185 | 0.47 |
| ISPH | 5,832 | 0.111 | 3135 | 2.03 |

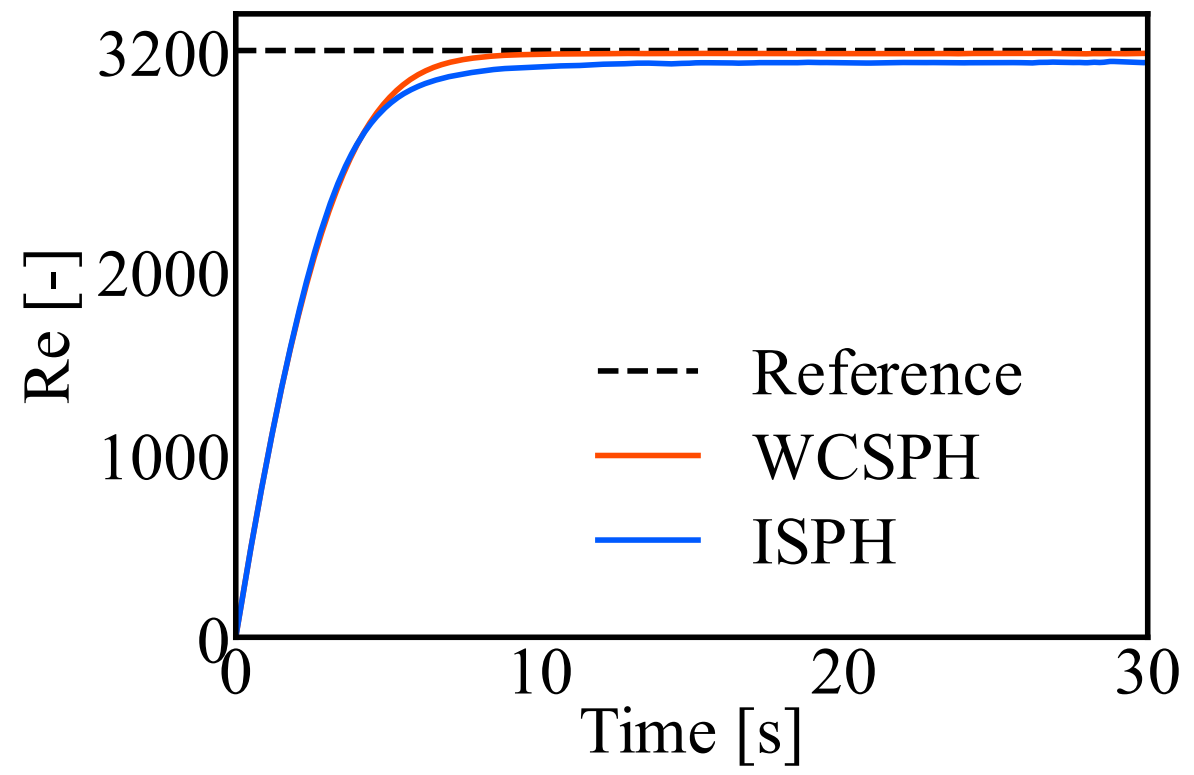

Fig. 13 Temporal history of predicted Re

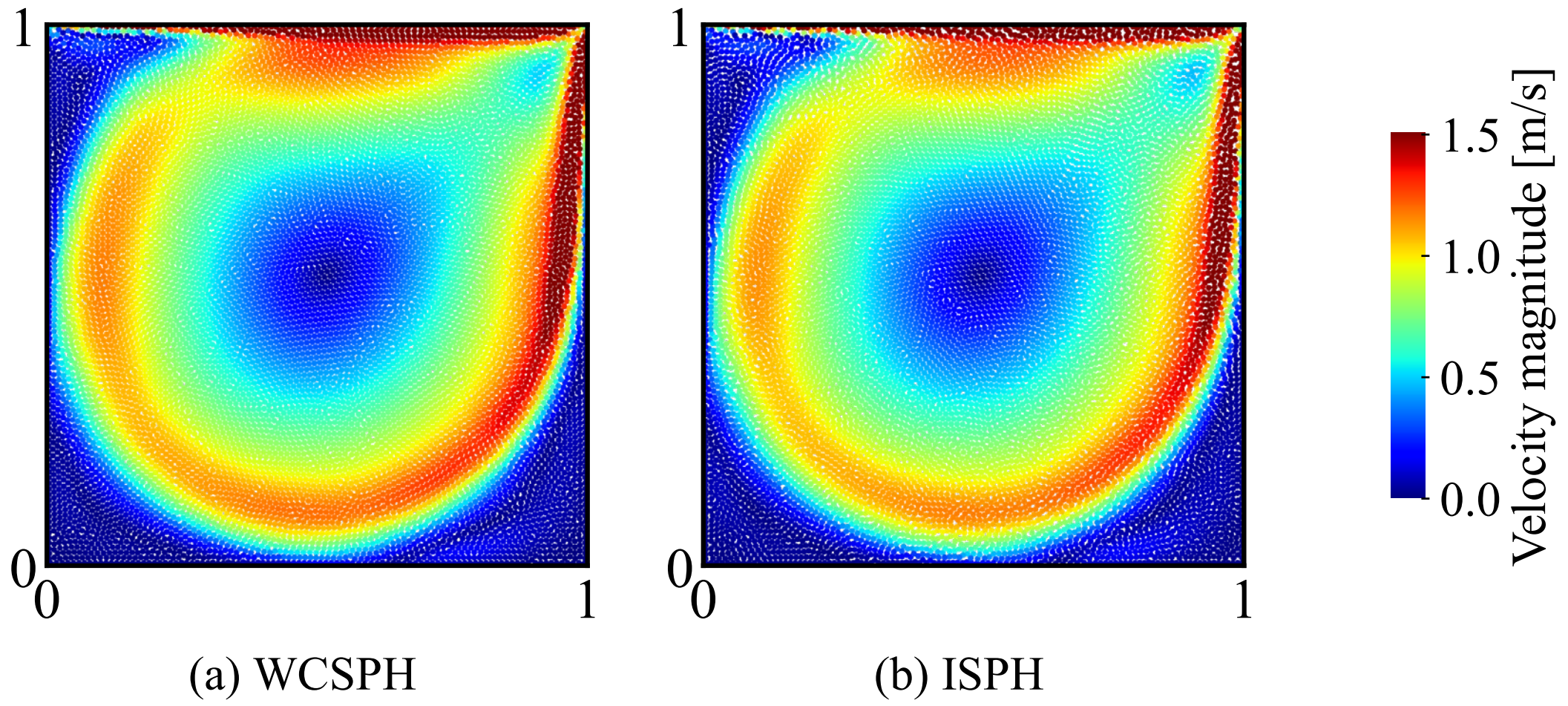

Fig. 14 Velocity field of each scheme at final time

Table 10 Comparison of loss gradients with respect to lid-velocity in the lid-driven cavity flow

| Scheme | FD | Prediction | Error [%] |
|---|---|---|---|
| WCSPH (AD) | $-1.8837 \times 10^{-1}$ | $-1.8837 \times 10^{-1}$ | $2.27 \times 10^{-11}$ |
| ISPH (AD) | $-1.8944 \times 10^{-1}$ | $-1.8946 \times 10^{-1}$ | $9.39 \times 10^{-3}$ |
| ISPH (AD+FD) | $-1.8944 \times 10^{-1}$ | $-1.8944 \times 10^{-1}$ | $9.25 \times 10^{-11}$ |

## 6. CONCLUSION

In this study, a composable and differentiable SPH framework, CraftSPH, has been developed. CraftSPH was designed as a modular framework rather than as a monolithic solver. Major numerical components are implemented as independent modules, while particle variables and material properties are managed through a unified particle object. For each problem, a solver is constructed by combining these components and specifying their computational sequence. This structure accommodates high-accuracy differential operators, stabilization techniques, boundary condition treatments, and matrix-form operators for semi-implicit pressure computation within the same framework, which also manages multiple particle types and fields. Individual components can also be extended when additional numerical functions are required. To examine this design, forward and inverse analyses were conducted using different combinations of the common components.

In the forward analyses, four fluid-flow problems were examined using ISPH with LSSPH. In Poiseuille flow, the analytical velocity profile was maintained despite the addition and removal of particles at the inlet and outlet. In the rising-bubble problem, a solver for two-fluid flow was constructed within the same framework, and the bubble shape, center-of-mass motion, and rise velocity agreed well with the reference solution. In the two- and three-dimensional dam-break problems, the pressure rises associated with fluid impact and subsequent variations were captured, although differences from the experimental measurements remained. In the three-dimensional dam-break analysis, the computational procedure developed for the two-dimensional problem was extended to a three-dimensional configuration, illustrating its reuse across dimensions. These results show that problem-specific solvers for the open-boundary, free-surface, two-fluid, and three-dimensional flow problems considered here can be constructed from common components.

CraftSPH was implemented entirely in PyTorch, enabling conventional SPH simulations and gradient computation based on AD to be performed within the same computational framework. In the inverse analyses, AD was applied to parameter estimation using both explicit WCSPH and semi-implicit ISPH with LSSPH. In the Taylor–Green vortex problem, the kinematic viscosity was estimated with errors of 0.19% and 12.13% for WCSPH and ISPH, respectively. In the lid-driven cavity problem, the Reynolds number was estimated with errors of 0.47% and 2.03%, respectively. These results demonstrate the application of AD to selected numerical operations in both schemes, while also indicating the importance of appropriately selecting the computational procedures included in the computational graph.

In conclusion, these analyses demonstrate that solvers assembled from common components can be applied to different forward problems and to parameter estimation with both WCSPH and ISPH. Together, these results demonstrate that users can configure the numerical procedure for a given problem and use the resulting solver for forward or inverse analysis, rather than being confined to a predetermined SPH solver. In particular, the ability to flexibly combine high-accuracy spatial discretization with semi-implicit pressure computation within the same framework, as demonstrated by the LSSPH-based

ISPH analyses, is a key contribution of this study. Several limitations of the present implementation should also be noted. The current validation mainly focuses on incompressible fluid flows, and support for more general higher-order differential operators, tensor-valued formulations, and coupling with other numerical methods remains to be further developed and validated. In addition, some discrete or iterative procedures are not always suitable for inclusion in automatic differentiation, and their exclusion may affect parameter-estimation accuracy. Future work will therefore focus on extending these capabilities and further integrating CraftSPH with deep-learning approaches.

## DECLARATION OF COMPETING INTERESTS

The authors declare that they have no known competing financial interests or personal relationships that could have appeared to influence the work reported in this paper.

## ACKNOWLEDGEMENTS

This work was supported by the JST BOOST (Grant Number: JPMJBS2414).

## DECLARATION OF GENERATIVE AI AND AI-ASSISTED TECHNOLOGIES IN THE MANUSCRIPT PREPARATION PROCESS

During the preparation of this work, the authors used ChatGPT by OpenAI to improve the language and clarity of the manuscript. After using this tool, the authors reviewed and edited the content as needed and took full responsibility for the content of the published article.